\documentclass[%
 reprint,
superscriptaddress,
 amsmath,amssymb,
 aps,
 prb,
 longbibliography,
]{revtex4-2}
\usepackage{CJK}
\usepackage{graphicx}
\usepackage{dcolumn}
\usepackage{amssymb}
\usepackage{amsbsy}
\usepackage{amsmath}
\usepackage{mathrsfs}
\usepackage{epsfig}
\usepackage{graphicx}
\usepackage{array}
\usepackage{textcomp}
\usepackage{color}
\usepackage{braket}
\usepackage{bm,hyperref}
\usepackage[justification=raggedright,font=small]{caption}
\usepackage[titletoc,toc,title]{appendix}
\usepackage[normalem]{ulem}
\usepackage[labelformat=simple]{subcaption}

\usepackage{tikz}

\usepackage{pgfplots}
\pgfplotsset{compat=1.18}
\usetikzlibrary{plotmarks}
\usetikzlibrary{external}
\tikzset{external/figure name = {fig}}

\usepackage{cleveref}
\crefname{appendix}{App.}{Apps.}
\crefname{equation}{Eq.}{Eqs.}
\crefname{figure}{Fig.}{Figs.}
\crefname{table}{Tab.}{Tabs.}
\crefname{section}{Sec.}{Secs.}

\newcommand{\beq}{\begin{equation}}
\newcommand{\eeq}{\end{equation}}

\newcommand{\bpm}{\begin{pmatrix}}
\newcommand{\epm}{\end{pmatrix}}
\newcommand{\bmm}{\begin{matrix}}
\newcommand{\emm}{\end{matrix}}

\usepackage{orcidlink}

\edef\flag{1}

\begin{document}
\preprint{APS/123-QED}

\title{Generalized Gibbs Ensemble from Eigenstate Entanglement Hamiltonian}

\author{Hao Chen\orcidlink{0000-0003-1084-9166}}
\affiliation{Department of Physics, Princeton University, Princeton, New Jersey 08544, USA}
\affiliation{Department of Electrical and Computer Engineering, Princeton University, Princeton, New Jersey 08544, USA}
\author{Biao Lian\orcidlink{0000-0002-8956-5619}}

\affiliation{Department of Physics, Princeton University, Princeton, New Jersey 08544, USA}

\date{\today}

\begin{abstract}
Relaxed quantum systems with conservation laws are believed to be approximated by the Generalized Gibbs Ensemble (GGE), which incorporates the constraints of certain conserved quantities serving as integrals of motion. By drawing an analogy between eigenstate reduced density matrix and GGE, we conjecture that a natural set of conserved quantities for GGE can emerge from the reduced density matrices of properly chosen eigenstates by the entanglement Hamiltonian superdensity matrix (EHSM) framework, and we demonstrate this explicitly for models mappable to free fermions. The framework proposes that such conserved quantities are linear superpositions of eigenstate entanglement Hamiltonians of a larger auxiliary system, where the eigenstates are Fock states occupying what we call the common eigenmodes, which remain eigenmodes when truncated within the physical subsystem. For 1D homogeneous free fermions with (anti-)periodic boundary conditions, which maps to 1D hardcore bosons with nearest neighbor hoppings, these conserved quantities lead to a non-Abelian GGE, which predicts the relaxation of both fermion and boson bilinears more accurately than the conventional Abelian GGE. Generalization of this framework may provide novel numerical insights for quantum integrability.

\end{abstract}

\maketitle

The dynamics of isolated quantum many-body systems has stimulated extensive interests in the past decades \cite{experiment_1,experiment_2,experiment_3,experiment_4,experiment_5,experiment_6,experiment_7}, which probes rich intrinsic properties of many-body systems related to and beyond thermalization, such as quantum chaos, integrability, many-body localization and quantum scars, etc. \cite{scar_1,scar_2,scar_3,EHSM31,EHSM32,EHSM33,EHSM34,EHSM35,EHSM36,EHSM37,EHSM38,EHSM39}. In thermalizing systems, the eigenstate thermalization hypothesis (ETH) \cite{ETH_srednicki1994, ETH_deutsch1991,conjecture2,conjecture3,conjecture4}  suggests that the reduced density matrix of a finite energy density eigenstate $\ket{\alpha}$ in a local subsystem $A$ resembles a thermal ensemble $\sim e^{-\beta H_A}$ where $H_A$ being the local Hamiltonian in $A$, giving rise to the picture that the only constraint on the system's dynamics is the local conservation of energy. In non-thermalizing quantum systems such as integrable models, the conservation laws generically lead to constrained quantum dynamics \cite{intro_1,intro_2,intro_3,intro_4,intro_5,intro_6} and hydrodynamics \cite{hydro_notes,hydro1,hydro2,hydro3,hydro4,hydro5,hydro6}. In particular, the generalized Gibbs ensemble (GGE) \cite{rigol_2007,rigol_2008,rigol_review} is introduced as a quantum statistical ensemble with the constraints of a certain set of conserved quantities (integrals of motion), providing a good approximation for evaluating the equilibrium values of observables in a quantum state after relaxing for long enough time.

However, it is intricate to determine which conserved quantities should be included in the GGE \cite{subtle1,subtle2,subtle3,subtle4,subtle5,subtle6,subtle7}. This issue is particularly acute in an integrable quantum system, where one expects the number of conserved quantities in the GGE to be proportional to the system size, thus they have to be selected from the exponentially numerous conserved quantities of the quantum model. In continuum quantum field theories (QFTs) where the concepts of integrability and locality have precise meanings, it is shown \cite{Essler_2015} that the optimal approach of matching the number of integrals of motion and the degrees of freedom is to include both (ultra-)local and quasi-local conserved operators in the GGE. However, systematically finding conserved quantities that constrain the relaxation dynamics for lattice systems remains an open question.

Recent studies have revealed that, as a natural generalization of ETH, the exponents of eigenstate reduced density matrices of non-thermalizing systems are approximately linear combinations of certain conserved operators \cite{conjecture1,biao_EHSM}, resembling the structure of GGE. In this letter, we conjecture that a natural GGE can be derived from the reduced density matrix of properly chosen eigenstates of a larger auxiliary system. We give the condition of such eigenstates for generic lattice models mappable to free fermions, and prove that the GGE emerged this way captures all the long-time averages of fermion bilinears exactly. Moreover, for the homogeneous 1D free fermion lattice model with (anti-)periodic boundary condition, which maps to the (interacting) hardcore boson model when there are only nearest neighbor hoppings, the GGE we arrive at is non-Abelian \cite{subtle7,NA1,NA2,NA3,NA4}, in contrast to the Abelian GGE initially proposed in \cite{rigol_2007}. We numerically verified that our non-Abelian GGE gives more accurate predictions of relaxations for the boson bilinear observables than the Abelian GGE.

\textit{The setup and conjecture.} Consider an isolated quantum many-body system $S$ with Hamiltonian $H_{S}$. The theory of GGE postulates that, for any initial state $|\Psi_S(0)\rangle$, the equilibrium value of a physical observable $O_S$ after a long-time relaxation is given by the average over the GGE density matrix $\rho_{\rm GGE}$:
 \beq
    \overline{\langle O_S\rangle}=\lim_{t\rightarrow\infty} \bra{\Psi_S(t)}O_S\ket{\Psi_{S}(t)} \approx {\rm Tr}(\rho_{\rm GGE}O_S), 
 \eeq
where $\ket{\Psi_S(t)} \equiv e^{-iH_St}\ket{\Psi_S(0)}$. The GGE density matrix $\rho_{\rm GGE}$ takes the form
\beq\label{eq:GGE}
\rho_{\rm GGE} = \frac{1}{Z}\exp\left(-\sum_{n=1}^{N_Q}\lambda_n Q_S^{(n)}\right)\ ,
\eeq
where $Q_S^{(n)}$ are a set of $N_Q$ conserved quantities (integrals of motion) commuting with $H_S$, and $\lambda_n$ are the Lagrange multipliers determined by the conserved expectation values  $\bra{\Psi_S(0)}Q_S^{(n)}\ket{\Psi_S(0)} = {\rm Tr}(\rho_{\rm GGE}Q_S^{(n)})$.

\begin{figure}
\centering

\begin{tabular}{ l l l }
    (a)&\ &(b)\\
\resizebox{0.75\columnwidth}{!}{%
    \if\flag1\includegraphics{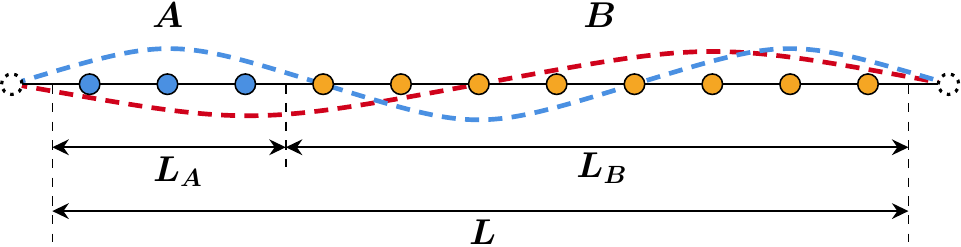}\else\include{fig1a}\fi
  }&\ &\resizebox{0.25\columnwidth}{!}{%
  \if\flag1\includegraphics{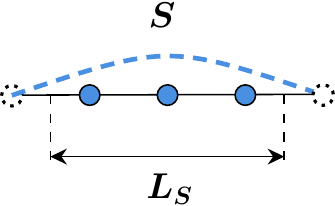}\else\include{fig1b}\fi
  }\\

    (c)&\ &(d)\\
\resizebox{0.75\columnwidth}{!}{%
    \if\flag1\includegraphics{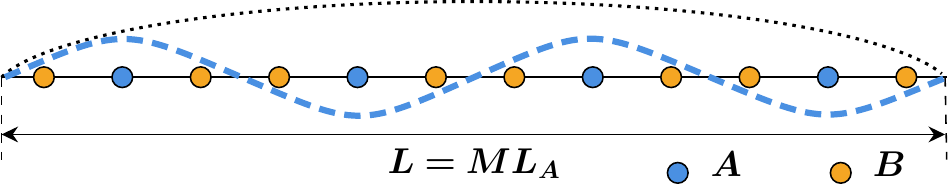}\else\include{fig1c}\fi
  }&\ &\resizebox{0.25\columnwidth}{!}{%
  \if\flag1\includegraphics{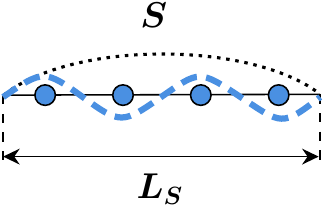}\else\include{fig1d}\fi
  }
\end{tabular}
\caption{
(a),(c) show the auxiliary system $A\cup B$ for homogeneous system $S$ in (b),(d) with OBC and PBC, respectively. Blue dashed curves (in (a),(c)) denote common eigenmodes in $A\cup B$ and their accordance in system $S$ (in (b),(d)). Red dashed curve in (a) is an eigenmode of $A\cup B$ which is not a common eigenmode.
}
    \label{fig1}
\end{figure}

It remains to determine a proper set of $N_Q$ conserved operators $Q_S^{(n)}$ for the GGE. While $Q_S^{(n)}$ could be chosen as the projectors onto eigenstates of $H_S$, this requires an exponentially large $N_Q$ equal to the dimension of the system's Hilbert space $\mathcal{H}_S$ \cite{rigol_2008}, which is significantly redundant.

We argue that $\rho_{\rm GGE}$, which imitates a relaxed state of isolated system $S$, resembles the reduced density matrix in a subsystem of some larger \emph{auxiliary system} in equilibrium.  
More specifically, we assume such an auxiliary system has a Hilbert space $\mathcal{H} = \mathcal{H}_A\otimes\mathcal{H}_B$ decomposable into two subsystems $A$ and $B$, where the subsystem $A$ have identical Hilbert space $\mathcal{H}_A\simeq \mathcal{H}_S$ and symmetries with the original isolated system $S$. Assume the auxiliary system $A\cup B$ has a Hamiltonian $H$, and is in an eigenstate $|\alpha\rangle$ of $H$ (thus in equilibrium). We define its reduced density matrix 
\begin{equation}\label{eq:rdmA}
\rho_A(\alpha)={\rm Tr}_B|\alpha\rangle\langle\alpha|=\exp\left(-H_E^A(\alpha)\right)\ , 
\end{equation}
and $H_E^A(\alpha)$ denotes the entanglement Hamiltonian of eigenstate $|\alpha\rangle$. As shown in \cite{biao_EHSM}, for eigenstates $|\alpha\rangle$ with finite energy densities, $H_E^A(\alpha)$ are approximately linear superpositions of conserved quantities of subsystem $A$ up to errors of boundary coupling terms between $A$ and $B$. Thus, $\rho_A(\alpha)$ of subsystem $A$ is analogous to $\rho_{\rm GGE}$ of system $S$.

The similarity between $\rho_A(\alpha)$ and $\rho_{\rm GGE}$ can be made exact by requiring the condition
\begin{equation}\label{eq-commu}
\left[H_E^A(\alpha),H_S\right]=0\ ,
\end{equation}
such that $H_E^A(\alpha)$ is conserved in system $S$ and resembles the exponent of GGE. Given a sufficiently large set of eigenstates $|\alpha\rangle$ satisfying \cref{eq-commu}, one can apply the EHSM method in \cite{biao_EHSM} (see SM. Sec. I \cite{SM}) to obtain a maximal set of orthogonal conserved operators $Q_A^{(n)}$ (which naturally map to operators $Q_S^{(n)}$ in system $S$) such that 
\begin{equation}\label{eq:EHSM-QS}
H_E^A(\alpha) = \sum_{n}\beta_A^{(n)}(\alpha)Q_A^{(n)}\simeq \sum_{n}\beta_A^{(n)}(\alpha)Q_S^{(n)}\ .
\end{equation}
In the resulting EHSM, each $Q_S^{(n)}$ has a weight $p_{A,n}\ge 0$, which is identified with the mean values of $\beta_A^{(n)}(\alpha)^2$ with respect to sufficiently many eigenstates $\ket{\alpha}$ (see SM \cite{SM} Sec. I for definition).

We conjecture that the set of conserved operators $Q_S^{(n)}$ with $p_{A,n}>0$ obtained from $\rho_A(\alpha)$ can be identified as conserved quantities contributing to $\rho_{\rm GGE}$ in \cref{eq:GGE}. Hereafter, we provide the exact condition for \eqref{eq-commu} to hold in models mappable to free fermion lattices, and show the $\rho_{\rm GGE}$ constructed describes the relaxation dynamics to a satisfactory extent. We expect the conjecture to also hold for generic interacting models if eigenstates satisfying \cref{eq-commu} can be found.

\emph{Free fermions.} Consider free fermion models (and equivalently models) with $L_S$ sites:
\begin{equation}\label{eq:Hfree}
H_S=\sum_{ij\in S}c^\dag_i h^S_{ij}c_j\ ,
\end{equation}
where $c^\dag_i$ and $c_i$ are the fermion creation and annihilation operators on site $i$, and $h^S_{ij}$ is the single-particle Hamiltonian. Assume the larger auxiliary system $A\cup B$ has $L$ sites and a Hamiltonian
\begin{equation}\label{eq:Haux-free}
H=\sum_{ij\in A\cup B}\tilde{c}^\dag_i h_{ij}\tilde{c}_j\ ,
\end{equation}
where $\tilde{c}_i$ and $\tilde{c}^\dag_i$ are the fermion creation and annihilation operators in the auxiliary system. We define a bijective function $g(\cdot)$ that maps each site $i$ in system $S$ to site $g(i)$ of subsystem $A$ (with number of sites $L_A=L_S$) of the auxiliary system $A\cup B$. We require subsystem $A$ and system $S$ to have the same symmetries.

For \cref{eq-commu} to hold, as we prove in SM.  Sec. II \cite{SM}, $H$ in \cref{eq:Haux-free} should be designed to have \emph{common eigenmodes} with $H_S$, and $|\alpha\rangle$ should be chosen as Fock states occupying the common eigenmodes of the auxiliary system. For each normalized single-particle eigenmode $\phi^S_{m,i}$ (creation operator $f^\dag_m$) of system $S$ ($1\le m\le L_S$) satisfying 
\begin{equation}\label{eq:mode-S}
\sum_{j\in S} h^S_{ij}\phi^S_{m,j}=\epsilon^S_m\phi^S_{m,i}\ ,\quad f^\dag_m=\sum_{i\in S} \phi^S_{m,i} c^\dag_i\ ,
\end{equation}
a \emph{common eigenmode} is defined as a normalized single-particle eigenmode $\phi^C_{m,i}$ of the auxiliary system (and its creation operator $\tilde{f}^\dag_m$) satisfying
\begin{equation}\label{eq:mode-C}
\begin{split}
&\sum_{j\in A\cup B} h_{ij}\phi^C_{m,j}=\epsilon_m\phi^C_{m,i}\ ,\quad \tilde{f}^{C\dag}_m=\sum_{i\in A\cup B} \phi^C_{m,i} \tilde{c}^\dag_i\ ,
\end{split}
\end{equation}
and is related to $\phi^S_{m,i}$ by
\begin{equation}\label{eq:common-eig}
\begin{split}
& \phi^C_{m,g(i)}=\mathcal{N}_m\phi^S_{m,i}\ ,\qquad (\forall i\in S)
\end{split}
\end{equation}
where $\mathcal{N}_m$ is a numerical factor independent of $i$. Namely, vector $\phi^S_{m,i}$ is the subset of components of $\phi^C_{m,i}$ in subsystem $A$. Note that $f^\dag_m$ in \cref{eq:mode-S} span a complete orthonormal eigenbasis of system $S$, while $\tilde{f}^{C\dag}_m$ in \cref{eq:mode-C} are only a subset of the eigenbasis of the auxiliary system $A\cup B$.

We then require the auxiliary system many-body eigenstates $|\alpha\rangle$ in \cref{eq:rdmA} to be Fock states occupying \emph{only} the common eigenmodes $\tilde{f}^{C\dag}_m$. Note that if there are energetically degenerate common eigenmodes, the eigenstates $|\alpha\rangle$ are allowed to occupy any modes which are their linear superpositions. Such Fock states optimally locally resemble equilibrium states, since eigenmodes at different energies in Fock states are uncorrelated, similar to equilibrium states upon time-averaging. The entanglement Hamiltonian of such Fock states takes the generic form of fermion bilinears \cite{Peschel_2003}:
\begin{equation}\label{eq:HEA-Fock}
\begin{split}
H_E^A(\alpha) &= \gamma(\alpha) I_A + \sum_{ij\in S}\kappa_{ij}(\alpha)\tilde{c}^\dag_{g(i)}\tilde{c}_{g(j)}\\
&\simeq \gamma(\alpha) I_S + \sum_{ij\in S}\kappa_{ij}(\alpha)c^\dag_{i}c_{j}\ ,
\end{split}
\end{equation}
where $\gamma(\alpha)$ is a constant, $I_A$ ($I_S$) is the identity operator in subsystem $A$ (system $S$), and $\kappa(\alpha)$ is a $L_S\times L_S$ Hermitian matrix. in the second line of \cref{eq:HEA-Fock}, the operator $H_E^A(\alpha)$ is mapped to an operator in system $S$ by mapping $\tilde{c}_{g(i)}\rightarrow c_i$ and $\tilde{c}^\dag_{g(i)}\rightarrow c^\dag_i$. It can then be proved (see SM. Sec. II, and Sec. III for a degenerate case \cite{SM}) that $H_E^A(\alpha)$ satisfies \cref{eq-commu}. As a direct result, from a sufficiently large number of such Fock states $|\alpha\rangle$, we can then obtain the set of conserved operators $Q_S^{(n)}$ (in \cref{eq:EHSM-QS}) by the EHSM method \cite{biao_EHSM}, which would be fermion bilinear operators (see SM. Sec. II for how these conserved operators depend on the choice of $\ket{\alpha}$ \cite{SM}), and construct $\rho_{\rm GGE}$ from them. Hereafter, we explicitly demonstrate the above idea for different 1D free fermion models.

\emph{Open boundary model.} We first consider system $S$ being a 1D homogeneous tight-binding model with open boundary condition (OBC), which has single-particle Hamiltonian $h^S_{ij}=-t_S(\delta_{i,j+1}+\delta_{i+1,j})$, 
where the sites range from $1\le i\le L_S$, and $-t_S$ is the real uniform nearest neighbor hopping. Its eigenmodes are sinusoidal standing waves (blue dashed curve in \cref{fig1}(b)):
\begin{equation}\label{eq:OBC-phiS}
\phi^S_{k,j}=\sqrt{\frac{2}{L_S+1}}\sin kj,\ \  (k=\frac{\pi m}{L_S+1},\ 1\le m\le L_S)
\end{equation}
with non-degenerate energies $\epsilon^S_k=-2t_S\cos k$.

The auxiliary system in this case can be chosen as a 1D tight-binding model in a chain of length $L=M(L_S+1)-1$ with OBC as shown in \cref{fig1}(a), with a single-particle Hamiltonian $h_{ij}=-t_0(\delta_{i,j+1}+\delta_{i+1,j})$, and $M>1$ is an integer. System $S$ maps to the subsystem $A$ of sites $1\le i\le L_S$ via the site map $g(i)=i$. The eigenmodes of the auxiliary system are sinusoidal waves $\phi_{\widetilde{k},j}=\sqrt{\frac{2}{L+1}}\sin \widetilde{k}j$, with $\widetilde{k}=\frac{\pi n}{L+1}$, $1\le n\le L$. In particular, when $n=Mm$, one has $\widetilde{k}=k$, and the eigenmode $\phi_{\widetilde{k},j}=\phi^C_{k,j}=\mathcal{N}_m\phi^S_{k,j}$ of the auxiliary system is a common eigenmode, which satisfies \cref{eq:common-eig} with factor $\mathcal{N}_m=\sqrt{\frac{L_S+1}{L+1}}=\frac{1}{\sqrt{M}}$. 

\begin{figure}
\centering
\begin{tabular}{ l l l }
    (a)&\ &(b)\\
    \if\flag1\includegraphics{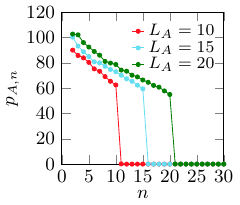}\else\include{essay_homo/fig2a}\fi&\ &\includegraphics[width=0.5\columnwidth]{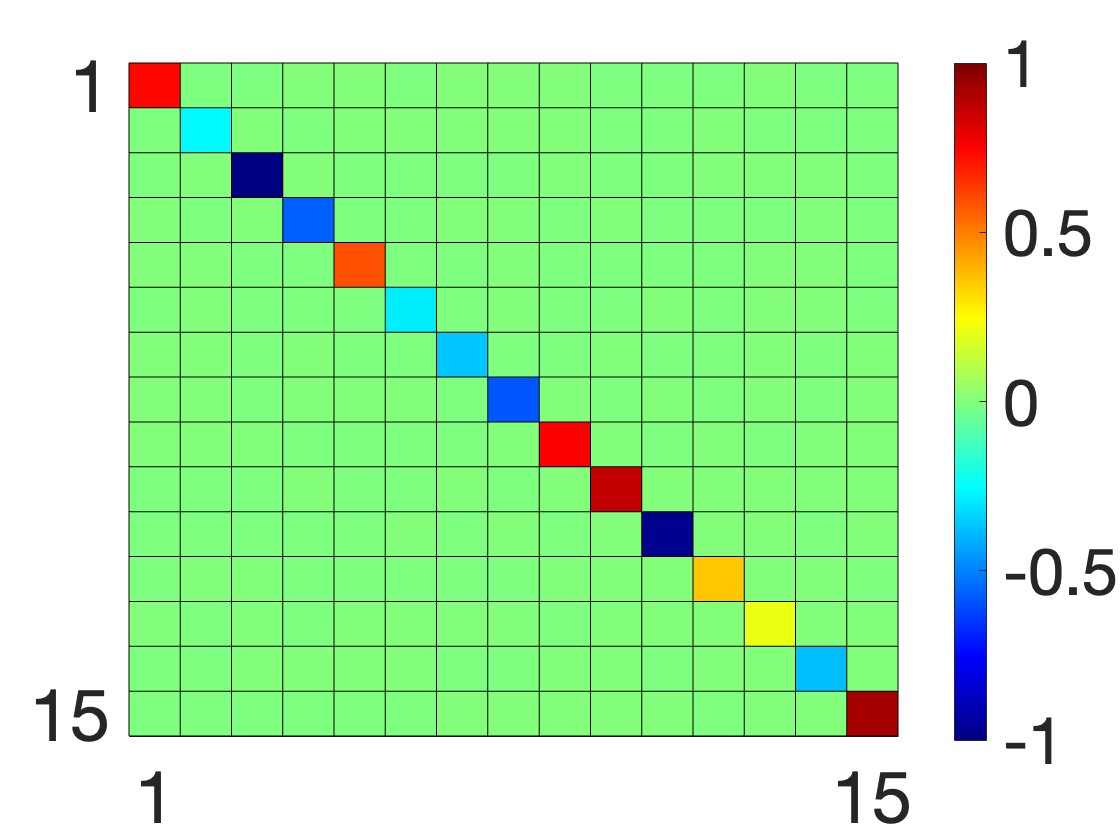}\\

    (c)&\ &(d)\\
    \if\flag1\includegraphics{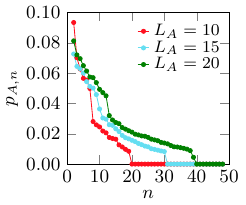}\else\include{fig2c}\fi&\ &\includegraphics[width=0.5\columnwidth]{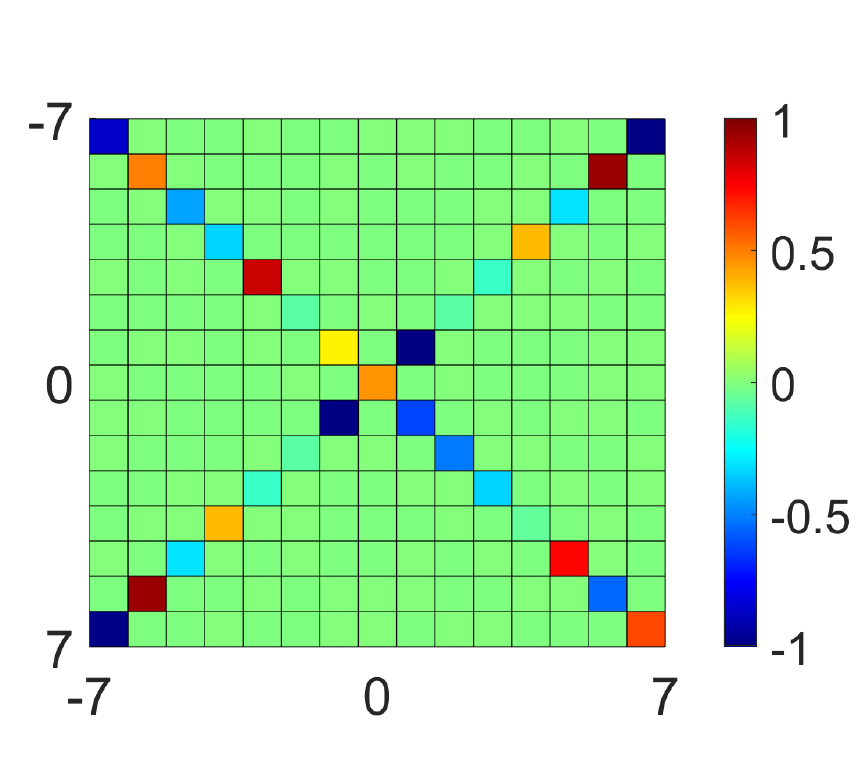}\\

    (e)&\ &(f)\\
    \includegraphics{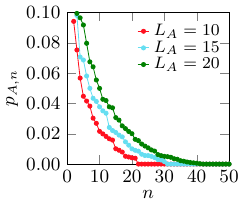}&\ &\includegraphics[width=0.5\columnwidth]{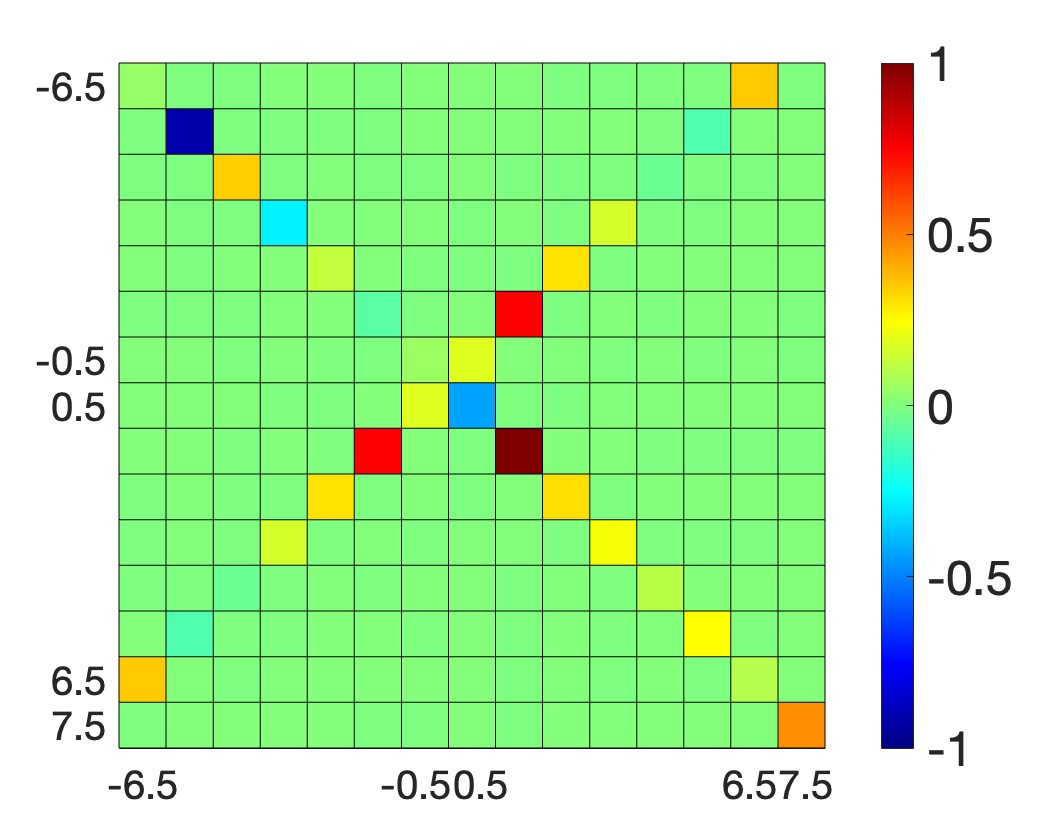}
\end{tabular}
\caption{
EHSM weights $p_{A,n}$ in descending order for 1D homogeneous free fermion chain with (a) OBC, (c) PBC, (e) A-PBC, in which $L=10,15,20$ and $L=20L_A$ respectively. (b) Matrix elements $\kappa_{Q,kk'}^{(n)}$ of a typical EHSM eigen-operator $Q_A^{(n)}\simeq Q_S^{(n)}=\sum_{k,k'}\kappa_{Q,kk'}^{(n)}f^\dag_k f_{k'}$ for OBC, with $k=\frac{\pi m}{L_S+1}$ in \cref{eq:OBC-phiS}. The horizontal and vertical labels are $1\le m\le L_S$. (d) and (f): Matrix elements $\kappa_{Q,kk'}^{(n)}$ of an EHSM eigen-operator $Q_A^{(n)}\simeq Q_S^{(n)}=\sum_{k,k'}\kappa_{Q,kk'}^{(n)}f^\dag_k f_{k'}$ for (d) PBC and (f) A-PBC, with $k=\frac{2\pi m}{L_S}$ in \cref{eq:PBC-phiS}. The horizontal and vertical labels are $-\frac{L_S}{2}< m\le \frac{L_S}{2}$. 
}
    \label{fig2}
\end{figure}

Applying the EHSM method \cite{biao_EHSM}, we find $L_S$ conserved operators $Q_S^{(n)}$ with EHSM weights $p_{A,n}>0$, as shown in \cref{fig2}(a) (sorted in descending order). Generically, we find these conserved operators $Q_S^{(n)}=\sum_{ij\in S}\kappa_{Q,ij}^{(n)}c^\dag_i c_j$ are linear superpositions of the conserved number operators (\cref{fig2}(b), and see SM. Sec. II for more details \cite{SM}) 
\begin{equation}
f^\dag_kf_k\ ,\qquad (k\in(0,\pi))\ ,
\end{equation}
where $f^\dag_k=\sum_{j\in S} \phi^S_{k,j} c^\dag_j$.

\emph{Periodic (Antiperiodic) boundary model.} We now turn to system $S$ being a 1D homogeneous tight-binding model in a length $L_S$ chain with periodic boundary condition (PBC) or antiperiodic boundary condition (A-PBC), and a real nearest neighbor hopping $-t_S$ (\cref{fig1}(d)). The eigenmodes are plane waves given by
\begin{equation}\label{eq:PBC-phiS}
\phi^S_{k,j}=\frac{e^{ikj}}{\sqrt{L_S}},\ \  (k=\frac{2\pi m}{L_S},\ -\frac{L_S}{2}< m\le \frac{L_S}{2})
\end{equation}
with energies $\epsilon_k^S=-2t_S\cos k$ which are 2-fold degenerate between $k$ and $-k$ (for $k\neq0$ or $\pi$), where for PBC (A-PBC), $m\in \mathbb{Z}$ ($m\in \mathbb{Z}+\frac{1}{2}$).

The corresponding auxiliary system $A\cup B$ is designed as a 1D tight-binding model in a length $L=ML_S$ chain with PBC (A-PBC) and real nearest neighbor hopping $-t_0$, where $M>1$ is an integer. The subsystem $A$ consists of sites which are integer multiples of $M$, and system $S$ maps to subsystem $A$ by function $g(j)=Mj$ ($1\le j\le L_S$). In this way, subsystem $A$ can be viewed as a \emph{coarse-grained} subregion of the auxiliary system, and preserves the translational symmetry (see \cref{fig1}(c)). Intriguingly, every eigenmode of the auxiliary system is a common eigenmode, which is a plane wave $\phi_{\widetilde{k},j}=\phi_{\widetilde{k},j}^C=\frac{1}{\sqrt{L}}e^{i\widetilde{k}j}$ with energy $\epsilon_{\widetilde{k}}=-2t_0\cos \widetilde{k}$, where $\widetilde{k}=\frac{2\pi n}{L}$, $-\frac{L}{2}<n\le \frac{L}{2}$. When restricted into subsystem $A$, one has $\phi_{\widetilde{k},g(j)}^C=\phi_{\widetilde{k},Mj}^C=\frac{1}{\sqrt{L}}e^{iM\widetilde{k}j}=\mathcal{N}_m\phi^S_{k,j}$, where $k=M\widetilde{k}\ (\text{mod }2\pi)$, $\mathcal{N}_m=\frac{1}{\sqrt{M}}$, satisfying \cref{eq:common-eig}. Thus, each eigenmode $\phi^S_{k,j}$ has $M$ common eigenmodes in the auxiliary system (see SM. Sec. III \cite{SM}).

By applying the EHSM method to generic auxiliary system Fock eigenstates $|\alpha\rangle$, which can occupy superposition of each pair of degenerate eigenmodes at momenta $\widetilde{k}$ and $-\widetilde{k}$, for PBC (A-PBC) we obtain $2L_S-\frac{3+(-1)^{L_S}}{2}$ ($2L_S-\frac{3-3(-1)^{L_S}}{2}$) linearly independent conserved quantities $Q_S^{(n)}$ with weight $p_{A,n}>0$, as shown in \cref{fig2}(c) (\cref{fig2}(e)). These $Q_S^{(n)}$ are found to be superpositions of the following conserved operators (\cref{fig2}(d,f), and see SM. Sec. III for details \cite{SM}):
\begin{equation}\label{eq:Qn-PBC}
f^\dag_k f_k,\ (k\in(-\pi,\pi])\ \ \  \& \ \ \  f^\dag_k f_{-k},\ (0<|k|<\pi)
\end{equation}
where $f^\dag_k=\sum_{j\in S} \phi^S_{k,j} c^\dag_j=\sum_{j}\frac{e^{ik j}}{\sqrt{L_S}}c^\dag_j$ is the momentum $k$ creation operator. In particular, $f^\dag_k f_{-k}$ is conserved since the energies of eigenmodes $k$ and $-k$ are degenerate. Note that $f^\dag_k f_{k}$ do not commute with $f^\dag_k f_{-k}$.

\begin{figure}
    \if\flag1\includegraphics{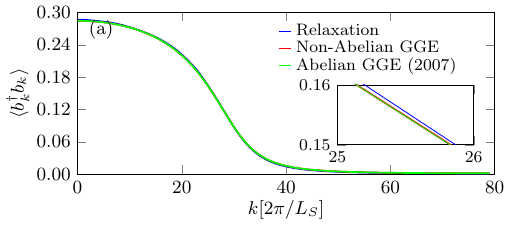}\\\else\include{essay_homo/fig3a}\fi
    \if\flag1\includegraphics{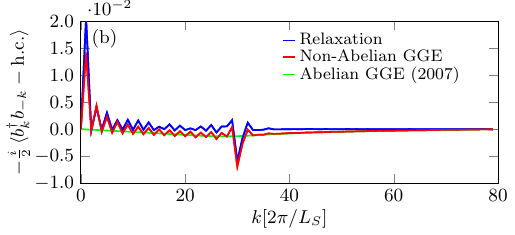}\else\include{essay_homo/fig3b}\fi
    
    \caption{The relaxed expectation values of hardcore boson bilinears of the initial state adopted in \cite{rigol_2007} (see description below \cref{eq-2p-fb}) with PBC, compared with ensemble averages from the Abelian and non-Abelian GGE. (a) The results for $\braket{b_k^\dagger b_k}$. (b) The results for $\text{Im}\langle b_{k}^\dag b_{-k}\rangle=-\frac{i}{2}\braket{b_k^\dagger b_{-k} - h.c.}$.}
    \label{fig3}
\end{figure}

\emph{Non-Abelian GGE.} In literature, it was proposed \cite{rigol_2007} that all conserved operators $Q_S^{(n)}$ in \cref{eq:GGE} commute with each other, which defines an \emph{Abelian GGE}. This well describes for example the OBC free fermion model (see SM \cite{SM} Sec. VI for numerical results). Here we consider a nontrivial example, the 1D hardcore boson model with PBC and nearest neighbor hoppings, which maps to the 1D free fermion model with PBC (A-PBC) for odd (even) particle number and uniform nearest hopping $-t_S$ here via the Jordan-Wigner transformation
\begin{equation}
a^\dag_j=\prod_{j'=1}^{j-1}(1-2c^\dag_{j'}c_{j'})c_j^\dag\ ,\qquad b^\dag_k=\sum_{j=1}^{L_S}\frac{e^{ik j}}{\sqrt{L_S}}a^\dag_j\ ,
\end{equation}
where $a^\dag_j$ and $b^\dag_k$ are the boson creation operators of site $j$ and momentum $k$, respectively. Ref. \cite{rigol_2007} proposed an Abelian GGE $\rho_\text{GGE}^{(A)}=Z^{-1}e^{-\sum_{k}\lambda_k f^\dag_k f_k}$ with $L_S$ commuting operators $f^\dag_k f_k$. Instead, \cref{eq:Qn-PBC} from our theory suggests a non-Abelian GGE $\rho_\text{GGE}^{(NA)}=Z^{-1}e^{-\sum_{k}(\lambda_k^{(+)} f^\dag_k f_{k}+\lambda_k^{(-)} f^\dag_k f_{-k})}$ with non-commuting conserved operators $f^\dag_k f_k$ and $f^\dag_k f_{-k}$, which was discussed in \cite{subtle7} but only for fermion $n$-body observables from a different motivation.

We examine whether our non-Abelian GGE predicts the relaxation well. First, it is straightforward to prove (see SM. Sec. VII \cite{SM}) that the long-time average of fermion bilinears are
\beq\label{eq-2p-fb}
    \overline{\braket{c_i^\dagger c_j}} = \sum_k \frac{e^{ik(j-i)}}{L_S}\braket{f_k^\dagger f_k} + \sum_{k\neq 0,\pi}\frac{e^{ik(j+i)}}{L_S}\braket{f_{-k}^\dagger f_{k}}
\eeq
because of the degeneracy between eigenmodes $f_k$ and $f_{-k}$. Thus, they can only be predicted accurately by our non-Abelian GGE $\rho_\text{GGE}^{(NA)}$, since the Abelian GGE $\rho_\text{GGE}^{(A)}$ yields $\braket{f_{-k}^\dagger f_{k}}=0$ for $k\neq 0,\pi$. 

Secondly, we examine the hardcore boson bilinears $\overline{\langle b_k^\dag b_{\pm k} \rangle}$, which are highly nonlocal, many-body operators in the fermion representation. We take the initial state adopted in \cite{rigol_2007}, which is $n=15$ particles in the ground state of a small box of length $L_0=40$ in the full chain of length $L_S=160$. We numerically calculate the long-time averages of hardcore boson two-point functions $\overline{\langle a_i^\dag a_j \rangle}$ via the method in \cite{HCB_two_point}, evaluate $\overline{\langle b_k^\dag b_{\pm k} \rangle}$ from them, and compare the results with the predictions from the two GGEs (see \cite{GGE_algo} and SM. Sec. IV for the algorithm evaluating boson two-point functions ${\rm Tr}[a_i^\dag a_j\rho_{\text{GGE}}]$ from GGE \cite{SM}). As shown in \cref{fig3}, our non-Abelian GGE predicts the relaxed values generically more accurately than the Abelian GGE in \cite{rigol_2007}. While the non-Abelian GGE predictions in this example are almost unchanged for $\langle b_k^\dag b_{k} \rangle$ (\cref{fig3}(a) inset), they are tremendously improved for $\langle b_{k}^\dag b_{-k}\rangle$  (see \cref{fig3}(b) and SM. Sec. VI \cite{SM}).

\begin{figure}
    \centering
    \includegraphics[width=\columnwidth]{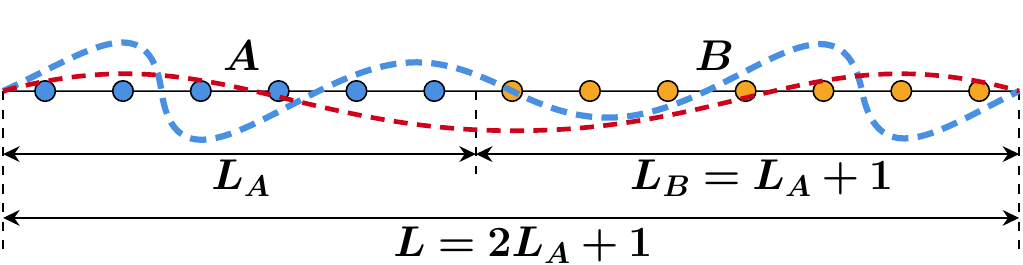}
    \caption{The mirror extension method of designing the auxiliary system for generic free fermion models. The blue (red) dashed curve represents an anti-symmetric (symmetric) eigenmode that is (is not) a common eigenmode.}
    \label{fig4}
\end{figure}
\emph{Generic models.} For generic free fermion models, we can design a mirror extension auxiliary system so that \cref{eq-commu} is satisfied for retrieving $\rho_{\rm GGE}$. For a generic free fermion system $S$ with $L_S$ sites and single-particle Hamiltonian $h^S_{ij}$, we can design an auxiliary system with $L=2L_S+1$ sites, where the single-particle Hamiltonian $h_{ij}=h_{L+1-i,L+1-j}$ is mirror symmetric about site $L_S+1$, and $h_{ij}=h^S_{ij}$ if $i,j\le L_S$. The subsystem $A$ of sites $1\le j\le L_S$ is identical to system $S$. All the auxiliary eigenmodes $\phi_{m,j}$ with mirror eigenvalue $-1$ (anti-symmetric) are common eigenmodes (\cref{fig4}) applicable for the EHSM method.

\emph{Discussion.} We demonstrated for free fermions that a natural GGE can be constructed by conserved quantities obtained from the reduced density matrices of proper eigenstates of a larger auxiliary system. For 1D free fermions with PBC (A-PBC) which have degenerate eigenmodes, the derived conserved operators do not commute, leading to a non-Abelian GGE, which predicts the relaxation of bilinear operators of fermions and hardcore bosons better than the conventional Abelian GGE \cite{rigol_2007}. It would be interesting to testify such a non-Abelian GGE in ultracold atoms. An important future question is to generalize this method for identifying GGE conserved quantities to interacting models, which requires designing auxiliary systems satisfying \cref{eq-commu} exactly or asymptotically. Potential ideas include weakly coupling the original system $S$ with an identical auxiliary system such that the eigenstates of the bipartite system are bonding/anti-bonding combinations of a system's eigenstate and its counterpart eigenstate of the auxiliary system, or by examining the mixed state entanglement Hamiltonian of eigenstates of nearby energies and minimizing the norm of $\left[H_E^A(\alpha),H_S\right]$. This may provide new insights into the numerical evidence for integrable quantum models.

\textit{Acknowledgements}. We are especially grateful to Fabian H. L. Essler and Wucheng Zhang for enlightening discussions that helped sharpen the idea of our paper. We thank Yumin Hu, Alex Jacoby, Abhinav Prem, Yifan (Frank) Zhang, Herman Verlinde, Jiabin Yu, Meng Cheng, Xueqi Chen, and Alisia Pan for their insightful comments. 
This work is supported by the Alfred P. Sloan Foundation, the National Science Foundation through Princeton University’s Materials Research Science and Engineering Center DMR-2011750, and the National Science Foundation under award DMR-2141966. Additional support is provided by the Gordon and Betty Moore Foundation through Grant GBMF8685 towards the Princeton theory program. H.C. receives additional supports from Bede Liu Fund for Excellence at the Department of Electrical and Computer Engineering of Princeton University.


\bibliography{arxiv2}

\newpage

\widetext
\clearpage
\begin{center}
\textbf{\large Supplemental Material}
\end{center}
\begin{center}
Hao Chen and Biao Lian \\
\today
\end{center}
\setcounter{equation}{0}
\setcounter{figure}{0}
\setcounter{table}{0}
\setcounter{page}{1}
\setcounter{section}{0}
\makeatletter
\renewcommand{\theequation}{S\arabic{equation}}
\renewcommand{\thefigure}{S\arabic{figure}}

\section{Review of the recipe of getting conserved quantities from entanglement Hamiltonian}\label{app:review}

This section is a recap of the EHSM method in \cite{biao_EHSM} on conserved quantities from entanglement Hamiltonian. We will only focus on the contents that are directly related to the purpose of this paper. 

The previous study \cite{biao_EHSM} was established on the fact that in a large system $A\cup B$ (with Hamiltonian $H$) which consists of two subregions: $A$ and its complement $B$ (FIG. 1a of the main text), for each many-body eigenstate $\ket{\alpha}$ (such that $H\ket{\alpha} = E_\alpha\ket{\alpha}$) of it, the corresponding entanglement Hamiltonian in subregion $A$: $H_E^A(\alpha)\equiv -\ln({\rm Tr}_B\ket{\alpha}\bra{\alpha})$ (where ${\rm Tr}_B\ket{\alpha}\bra{\alpha}\equiv\rho_A(\alpha)$ is the reduced density matrix) is a linear combination of a set of linearly independent conserved operators $\{Q_A^{(n)}\}, (n=0,1,2...)$ that are \textit{localized in subregion $A$}:
\beq
    H_E^A(\alpha) = \sum_{n}\beta_A^{(n)}(\alpha)Q_A^{(n)},
\eeq
where $\beta_A^{(n)}(\alpha)$ are just the coefficients specified by $\ket{\alpha}$. By saying an operator is ``localized in subregion $A$'' we mean the support of its each product term is in subregion $A$ but can extend to the entire subregion instead of being an ``local operator'' in the common sense. The rationale underlying this fact is that the reduced density matrix $\rho_A(\alpha)$ is a conserved operator under the time evolution governed by $H$: $\rho_A(\alpha,t) = {\rm Tr}_B \left(e^{-iE_\alpha t}\ket{\alpha}\bra{\alpha}e^{iE_\alpha t}\right) =\rho_A(\alpha)$, so as its logarithm $H_E^A(\alpha)$. 

Therefore, with sufficiently many $H_E^A(\alpha)$'s, we can reconstruct the maximal set of linearly independent conserved operators $\{Q_A^{(n)}\}$. The algorithm, called EHSM, was presented in \cite{biao_EHSM} and is briefly reviewed here. Taking suffiiently many eigenstates $\ket{\alpha}$'s (forming an ensemble $\Xi$), the corresponding $H_E^A(\alpha)$'s lie in a linear space of $N_A\times N_A$ matrices, where $N_A$ is the Hilbert space dimension of subsystem $A$. This indicates that one can reconstruct the independent $Q_A^{(n)}$'s by finding a basis of this space. The numerical algorithm of finding a basis involves constructing a \textit{entanglement Hamiltonian super-density matrix} (EHSM) and diagonalizing it: 
\begin{equation}
R_A \equiv \frac{1}{N_A}\sum_{\alpha\in\Xi}\left|H_E^A(\alpha)\right)\left(H_E^A(\alpha)\right| = \sum_{n}p_{A,n}\left|\overline{Q}_A^{(n)}\right)\left(\overline{Q}_A^{(n)}\right|\ , 
\end{equation}
where $p_{A,n}\ge 0$ is the $n$th eigenvalue of $R_A$ in descending order, and $\overline{Q}_A^{(n)}$ is the orthonormalized eigen-operator satisfying $\left(\overline{Q}_A^{(m)}|\overline{Q}_A^{(n)}\right)=\text{tr}\left(\overline{Q}_A^{(m)\dag}\overline{Q}_A^{(n)}\right)=\delta_{mn}$, which resembles subregionally quasi-local conserved quantities in subregion $A$ if its eigenvalue $p_{A,n}>0$.

Generically, $R_A$ is an operator with huge dimensions $N_A$ that is hard to diagonalize when doing numerical calculations. However, one only needs to diagonalize a much smaller correlation matrix (matrix size given by the number of eigenstates $|\alpha\rangle$ used)
\begin{equation}
K_{A,\alpha\beta}\equiv\frac{1}{N_A}\left(H_E^A(\alpha)|H_E^A(\beta)\right)\ ,
\end{equation}
which has the same nonzero eigenvalues $p_{A,n}>0$, and the corresponding conserved operators $\left|\overline{Q}_A^{(n)}\right)$ can be derived from the eigenvectors of $K_{A,\alpha\beta}$, as proved in \cite{biao_EHSM}. 

Based on the fact that $Q_A^{(n)}$ are conserved, we formally stated a conjecture \cite{conjecture1,conjecture2,conjecture3,conjecture4} that these operators $Q_A^{(n)}$ are not only conserved in the large system, but also resemble conserved operators $Q_S^{(n)}\simeq Q_A^{(n)}$ in a small isolated system $S$ (with its own Hamiltonian $H_S$) that has a similar physical structure as subregion $A$, or in mathematical language: $[H_S,Q_S^{(n)}]=0$. Then as long as this conjecture holds, we can use the EHSM algorithm to find a set of conserved quantities of the small system $S$ by taking enough many eigenstates $\ket{\alpha}$'s of the large system. 

Unfortunately, it is readily apparent that this conjecture holds if and only if $[H_S,H_E^A(\alpha)]=0$ for \textit{every} eigenstate $\ket{\alpha}$ of the large system, which is generally not true given an arbitrary $H_S$. If we impose the EHSM algorithm anyway, it returns a mixture of conserved and non-conserved operators in the system $S$. However, we can reverse the way of asking the question: given the geometry of a small isolated system $S$ and its Hamiltonian $H_S$, how should we design a larger auxiliary system, making $S$ to have a similar physical structure of a subregion $A$ of it for the conjecture to hold, in order to find conserved operators of $S$ using the EHSM algorithm in \cite{biao_EHSM}? This question is difficult to answer in generic situations, which we leave for future works, but we find neat results for non-interacting fermion models, as discussed in the main text. 


\section{Conservation of entanglement Hamiltonian in presence of common eigenmodes}\label{app:conserve_EH}
In this section, we prove that for free fermion models, the entanglement Hamiltonian of an eigenstate $\ket{\alpha}$ in the auxiliary system, with only common eigenmodes occupied, is conserved in the isolated system $S$, which is the key factor for the EHSM algorithm to give conserved quantities in $S$. This proof relies on the method of calculating reduced density matrices from two-point
correlation functions for non-interacting systems by Peschel \cite{Peschel_2003} and the existence of eigenmodes. 

Before doing any proof, we need to specify the relationship between system $S$ and the auxiliary system $A\cup B$, as we did in the main text. First of all, there is a \emph{bijective} (surjective and injective) map $g(\cdot)$ between sites in $S$ and the sites in subregion $A$ of the auxiliary system such that given any site $i$ in $S$, there is a unique corresponding site $g(i)\in A$ in the auxiliary system. For instance, in the homogeneous OBC cases, we discussed in the main text, $g(i) = i$; in the PBC case with coarse-grained subregion, $g(i) = Mi$. 

Second, we want there to be a set of \emph{common eigenmodes} between the two systems, which resembles all the eigenmodes in $S$. This means that for any eigemode $\phi_{m,i}^S$ of $S$, there exists an eigenmode $\phi_{m,j}^C$ (or more than one eigenmodes, which we will discuss an example in the next section) of the auxiliary system such that
\beq
    \forall i\in S,\ \ \phi_{m,g(i)}^C = \mathcal{N}_m\phi_{m,i}^S,
\eeq
which means that if we only look at the sites $j=g(i)$ that corresponds to all sites $i$'s in the system $S$, the eigenmode wavefunction $\phi_{m,g(i)}^C$ of the auxiliary system has the same shape as $\phi_{m,i}^S$.

Third, system $S$ has its own creation and annihilation operators: $c_i^\dagger, c_i,(i\in S)$, while the auxiliary system has its own: $\tilde{c}_j^\dagger, \tilde{c}_j,(j\in A\cup B)$. The two systems become related once we assign $\tilde{c}_{g(i)} \rightarrow c_i$ for all $i\in S$. Then we are enabled to use conserved operators in the subregion $A$ of the auxiliary system to resemble conserved operators in system $S$.

In reference \cite{Peschel_2003}, it is shown that for free-fermion systems, the entanglement Hamiltonian of any \textit{Fock state} $\ket{\Psi}$ is closely related to the $L\times L$ correlation matrix (two-point functions)
\beq
    \mathcal{C}_{ij}(\Psi) = \braket{\tilde{c}_i^\dagger \tilde{c}_j}\equiv \bra{\Psi}\tilde{c}_i^\dagger \tilde{c}_j\ket{\Psi},
\eeq
where $i,j\in A\cup B$. Additionally, Wick's theorem tells us that any four-point or higher-order function can be expressed by $\mathcal{C}$ via pairing the operators like
\beq
    \braket{\tilde{c}_i^\dagger \tilde{c}_j^\dagger \tilde{c}_m \tilde{c}_n} = \braket{\tilde{c}_i^\dagger \tilde{c}_n}\braket{\tilde{c}_j^\dagger \tilde{c}_m}-\braket{\tilde{c}_i^\dagger \tilde{c}_m}\braket{\tilde{c}_j^\dagger \tilde{c}_n}.
\eeq

By definition, the reduced density matrix $\rho_A(\Psi) = e^{-H_E^A(\Psi)}$ must yield correct correlation functions at any order, which is equivalent to the following two properties:
\begin{enumerate}
    \item Correct two-point functions $\braket{c_i^\dagger c_j}$ for any two sites $i,j\in$ subregion $A$
    \beq
        \mathcal{C}_{A,ij}(\Psi)\equiv{\rm Tr}\left(\rho_A(\Psi)\tilde{c}_i^\dag \tilde{c}_j\right) = \braket{\tilde{c}_i^\dagger \tilde{c}_j}, \ \ \ i,j\in A,
    \eeq
    where $\mathcal{C}_A$ stands for the $L_A\times L_A$ (sub-)correlation matrix in $A$.
    \item Any four-point or higher-order correlation functions should be related to two-point functions in the same way as given by Wick's theorem, for example, $i,j,m,n\in A$:
    \beq
    \begin{aligned}
        &{\rm Tr}\left(\rho_A(\Psi)\tilde{c}_i^\dagger \tilde{c}_j^\dagger \tilde{c}_m \tilde{c}_n\right) \\
        = &{\rm Tr}\left(\rho_A(\Psi)\tilde{c}_i^\dag \tilde{c}_n\right){\rm Tr}\left(\rho_A(\Psi)\tilde{c}_j^\dag \tilde{c}_m\right) \\
        &- {\rm Tr}\left(\rho_A(\Psi)\tilde{c}_i^\dag \tilde{c}_m\right){\rm Tr}\left(\rho_A(\Psi)\tilde{c}_j^\dag \tilde{c}_n\right)
    \end{aligned}
    \eeq
\end{enumerate}

It can be shown that the second property holds if $\rho_A(\Psi)$ is the exponential of a bilinear operator, which means
\beq
    \rho_A(\Psi) = \mathcal{K}\exp\left(\sum_{i,j\in A}\kappa_{A,ij}(\Psi)\tilde{c}_i^\dagger \tilde{c}_j\right),
\eeq
where $\mathcal{K}^{-1}={\rm Tr}\exp(...)$ is the normalization factor to make the partial trace over subregion $A$ of $\rho_A(\Psi)$ unity. Then, to carry out the first property above, we find the $L_A\times L_A$ matrix $\kappa_A(\Psi)$ should have the same set of eigenvectors $v_{s,j}$ (here $s$ denotes a set of suitable indices for these eigenvectors) as the transpose of correlation matrix $\mathcal{C}_A$:
\beq\label{eq:ckev}
\begin{aligned}
    \kappa_{A,ij}(\Psi) &= \sum_{s}v_{s,i}v_{s,j}^*\varepsilon_s,\\
    \mathcal{C}_{A,ij}(\Psi) &= \sum_{s}v_{s,i}^*v_{s,j}\zeta_s.
\end{aligned}
\eeq
where $\varepsilon_s$ and $\zeta_s=(e^{\varepsilon_s}+1)^{-1}$ are their eigenvalues of mode $s$. In brief, $\kappa^T = \ln[(\mathcal{C}_A^{-1}-I)]$ where $I$ is the $L_A\times L_A$ identity matrix. 

The primary result from the above derivation is that the entanglement Hamiltonian is directly given by the correlation matrix $\mathcal{C}_A$:
\beq
    H_E^A(\Psi) = \gamma(\Psi)I_A + \sum_{i,j\in A}\kappa_{A,ij}(\Psi)\tilde{c}_i^\dagger \tilde{c}_j,
\eeq
where $\gamma(\Psi) = \ln(1/\mathcal{K})= -{\rm Tr}\ln(I-\mathcal{C}_A(\Psi))$ can be worked out via direct calculation. 

As we mentioned in the main text, by relating the operators in subregion $A$ and those in system $S$ via $\tilde{c}_{g(i)} \rightarrow c_i$ for all sites $i\in S$, we can rewrite the above expression using site indices in system $S$
\beq\label{eq:EH_corr}
    H_E^A(\Psi) = \gamma(\Psi)I_A + \sum_{i,j\in S}\kappa_{A,g(i)g(j)}(\Psi)\tilde{c}_{g(i)}^\dagger \tilde{c}_{g(j)} \simeq \gamma(\Psi)I_S + \sum_{i,j\in S}\kappa_{ij}(\Psi)c_i^\dagger c_j ,
\eeq
where matrix $\kappa(\Psi)$ is defined by matching its elements with matrix $\kappa_A(\Psi)$: $\kappa_{ij}(\Psi)\equiv \kappa_{A,g(i)g(j)}(\Psi)$. To see whether $H_E^A(\alpha)$ for auxiliary system eigenstate $\ket{\alpha}$ is conserved under the evolution governed by $H_S$,  which can be seen from whether $[H_S,H_E^A(\alpha)] = 0$, 
we only need to examine if $H_S$ and $H_E^A(\alpha)$ can be simultaneously diagonalized.

We take the state $\ket{\Psi}$ of the large system to be a many-body eigenstate $\ket{\alpha}$ that is generally (here is a subtlety if the system has degenerate modes, which we will discuss in next section) given by
\beq\label{eq:eigenstate}
    \ket{\alpha} = \prod_{l=1}^L \left(\tilde{f}_l^\dagger\right)^{\eta_{\alpha,l}}\ket{0},
\eeq
where $\tilde{f}_l^\dagger = \sum_{j\in A\cup B} \phi_{l,j}c_j^\dagger$ is the creation operator for the $l$-th eigenmode of the auxiliary system, and $\eta_{\alpha,l} = 0,1$ is the corresponding occupation number. Here we have used $\phi_{l,j}$ to represent the (normalized) $l$th single-particle eigenmode wavefunction, which diagonalizes the auxiliary system Hamiltonian in the following way
\beq
\begin{aligned}
    H &\equiv \sum_{ij\in A\cup B}h_{ij}\tilde{c}_i^\dag \tilde{c}_j = \sum_{ij}h_{ij}\left(\sum_l\phi_{l,i}^*\tilde{f}_l^\dag\right)\left(\sum_{l'}\phi_{l',j}\tilde{f}_{l'}\right) = \sum_{ll'}\tilde{f}_l^\dag \tilde{f}_{l'}\sum_{ij}h_{ij}\phi_{l,i}^*\phi_{l',j}= \sum_{ll'}\tilde{f}_l^\dag \tilde{f}_{l'}\epsilon_l\delta_{ll'}\\
    &= \sum_{ij}\tilde{c}_i^\dag \tilde{c}_j\sum_{l}\epsilon_l\phi_{l,i}\phi_{l,j}^*\Longrightarrow h_{ij} = \sum_{l}\epsilon_l\phi_{l,i}\phi_{l,j}^*.
\end{aligned}
\eeq
Here we remind that in the complete set of eigenmodes $\{\phi_{l,i}|1\leq l\leq L\}$, there is a subset of common eigenmodes $\{\phi_{l_m,i}|\phi_{l_m,i} = \phi_{m,i}^C, 1\leq m\leq L_S\}$. 

Then, it can be verified that the correlation matrix in the entire auxiliary system is
\beq
    \mathcal{C}_{ij}(\alpha) = \bra{\alpha}\tilde{c}_i^\dagger \tilde{c}_j\ket{\alpha} = \sum_{l=1}^L \eta_{\alpha,l}\phi_{l,i}^*\phi_{l,j},\ \ \ i,j\in A\cup B, 
\eeq
which leads to the correlation matrix in the subregion $A$
\beq
    \mathcal{C}_{A,ij}(\alpha) =  \sum_{l=1}^L \eta_{\alpha,l}\phi_{l,i}^*\phi_{l,j},\ \ \ i,j\in A,
\eeq
or equivalently, writing in terms of the site indices in system $S$
\beq
    \mathcal{C}_{A,g(i)g(j)}(\alpha) =  \sum_{l=1}^L \eta_{\alpha,l}\phi_{l,g(i)}^*\phi_{l,g(j)},\ \ \ i,j\in S.
\eeq

Now, we impose the restriction that $\ket{\alpha}$ can only have common eigenmodes occupied, i.e. $\eta_{\alpha,l}=0$ for any eigenmode $l$ that is not common between the two systems. As a result, the existence of common eigenmodes $\phi_{l_m,g(i)} = \phi_{m,g(i)}^C = \mathcal{N}_m\phi_{m,i}^S$ helps us to rewrite 
\beq\label{eq:derive_corr}
    \begin{aligned}
    \mathcal{C}_{A,g(i)g(j)}(\alpha) &= \sum_{l_m\in\text{common}} \eta_{\alpha,l_m}\phi_{l_m,g(i)}^*\phi_{l_m,g(j)}\\
    &= \sum_{m=1}^{L_S} \eta_{\alpha,l_m}\phi_{m,g(i)}^{C*}\phi_{m,g(j)}^C\\
    &= \sum_{m} |\mathcal{N}_m|^2\eta_{\alpha,l_m}\phi_{m,i}^{S*}\phi_{m,j}^S, \ \ \ i,j\in S
    \end{aligned}
\eeq
This result shows that the common eigenvectors $v_{s,i}$ of $\mathcal{C}_A(\alpha)$ and $\kappa_A(\alpha)$ in (\ref{eq:ckev}) are just the eigenmode wavefunctions of $H_S$, indicating that the entanglement Hamiltonian (\ref{eq:EH_corr}) can be diagonalized simultaneously with $H_S$ and thus $\left[H_S,H_E^A(\alpha)\right] = 0$:
\beq
\begin{aligned}
    &\kappa_{ij}(\alpha) = \kappa_{A,g(i)g(j)} = \sum_{m=1}^{L_S}\varepsilon_{m}\phi_{m,i}^S\phi_{m,j}^{S*},\ \ \  \varepsilon_m = \ln\left[\frac{1}{|\mathcal{N}_m|^2\eta_{\alpha,l_m}}-1\right]\\
    &H_S = \sum_{ij\in S}h_{ij}^S c_i^\dag c_j, \ \ \ h_{ij}^S = \sum_{m=1}^{L_S}\epsilon_m^S\phi_{m,i}^S\phi_{m,j}^{S*}
\end{aligned}
\eeq

More importantly, one can read from \cref{eq:EH_corr} that $H_E^A(\alpha)$ is a linear combination of $f_m^\dag f_m$, which proves that the eigen-operators $\overline{Q}_A^{(n)}$ with $p_{A,n}>0$ of the EHSM $R_A$ over sufficiently many auxiliary system eigenstates form a complete basis of the linear space spanned by $f_m^\dag f_m$ for all the eigenmodes of system $S$. Therefore, the EHSM method can successfully capture all the bilinear conserved operators.

It is evident that the above derivation is generic as long as the existence of common eigenmodes is assumed. Therefore, we can argue that: for any given isolated free-fermion lattice system $S$ whose Hamiltonian is $H_S$, we can design an auxiliary system with a Hamiltonian $H$ such that:
\begin{enumerate}
    \item There is a subregion $A$ (not necessarily connected) of the auxiliary system that has the same (geometric) structure as $S$, which we will investigate its entanglement with the rest part of the auxiliary system.
    \item \textit{A subset of} the eigenmodes of $H$ resembles \textit{all the} eigenmodes of $H_S$ if we only look at the sites in subsystem $A$. 
\end{enumerate}
Then, for any eigenstate $\ket{\alpha}$ whose occupied eigenmodes belong to the subset mentioned above, its corresponding entanglement Hamiltonian $H_E^A(\alpha)$, which is a linear combination of the mode occupation operators $f_m^\dag f_m$ of the common eigenmodes, commutes with $H_S$. A practical design for generic OBC systems was given in the main text.

\section{Conservation of entanglement Hamiltonian in coarse-grained subregion}\label{app:coarse_grain}
Most of the proof in the above section can be directly applied to the PBC example with the coarse-grained subregion we discussed in the main text, except for the common eigenmodes between two systems are complicated: one eigenmode in $S$ has $M$ corresponding eigenmodes in the auxiliary system. 

For the coarse-grained setup we had in the main text, the sites in $S$ and the auxiliary system match in the following way: As the sites in system $S$ are indexed by integers $1\leq j\leq L_S$, while the sites in the auxiliary system $A\cup B$ are indexed by integers $1\leq j\leq L$, site $j$ in $S$ is mapped to site $g(j) = Mj$ in the auxiliary system. The crucial point is that every eigenmode wavefunction of the auxiliary system $\phi_{\widetilde{k},j} = \frac{1}{\sqrt{L}}e^{i\widetilde{k}j}$, where $\widetilde{k} = \frac{2\pi n}{L}$, $-\frac{L}{2}<n\le \frac{L}{2}$ is the quasi-momentum, is a common eigenmode. In this situation, we use $\widetilde{k}$ to index all the common eigenmodes, such that 
\beq
    \phi_{\widetilde{k},j}^C=\phi_{\widetilde{k},j}=\frac{1}{\sqrt{L}}e^{i\widetilde{k}j}.
\eeq
Knowing the eigenmodes of $S$ take the similar form of plane waves: $\phi^S_{k,j} = \frac{1}{\sqrt{L_S}}e^{ikj}$, $k = \frac{2\pi m}{L_S}$, $-\frac{L_S}{2}<m\le \frac{L_S}{2}$, it can be seen that when restricted to the coarse-grained subregion $A$, the common eigenmode $\phi_{\widetilde{k},g(j)}^C$ corresponds to an eigenmode in $S$:
\beq
    \phi_{\widetilde{k},g(j)}^C=\phi_{\widetilde{k},Mj}^C=\frac{1}{\sqrt{L}}e^{iM\widetilde{k}j}=\mathcal{N}_m\phi^S_{k,j},\ \ \ j\in S,
\eeq
where $k=M\widetilde{k}\ (\text{mod }2\pi)$, $\mathcal{N}_m=\frac{1}{\sqrt{M}}$. It can be seen from here that each eigenmode $\phi^S_{k,j}$ corresponds to $M$ eigenmodes in the auxiliary system.

As a result, the correlation matrix derived in (\ref{eq:derive_corr}) becomes (notation 1BZ denotes the first Brillouin zone, which is the interval $(-\pi,\pi]$)
\beq\label{eq:corr_degenerate}
    \begin{aligned}
    \mathcal{C}_{A,g(i)g(j)}(\alpha) &= \frac{1}{L}\sum_{\widetilde{k}\in 1\text{BZ}}\eta_{\alpha,\widetilde{k}} e^{-i\tilde{k}(Mi-Mj)}\\
    &= \frac{1}{L}\sum_{\widetilde{k}\in 1\text{BZ}}\eta_{\alpha,\widetilde{k}} e^{-ik(i-j)},\ \ (k\equiv M\widetilde{k}, \text{do not take mod $2\pi$ for the moment})\\
    &= \frac{1}{L}\sum_{n}\eta_{\alpha,n}e^{-ik(i-j)},\ \ (-\frac{L}{2}<n\le\frac{L}{2},\ \eta_{\alpha,n}\equiv\eta_{\alpha,\widetilde{k}},\ \widetilde{k} = \frac{2\pi n}{L},\ k = \frac{2\pi n}{L_S})\\
    &= \frac{L_S}{L}\sum_{n}\eta_{\alpha,n}\phi_{n,i}^{S*}\phi_{n,j}^S,\ \ (\phi_{n,j}^S\equiv \frac{1}{\sqrt{L_S}}e^{ikj}, \ \text{in which } k = \frac{2\pi n}{L_S})\\
    &= \frac{L_S}{L}\sum_{m}\sum_{N}\eta_{\alpha,(m+NL_A)}\phi_{k,i}^{S*}\phi_{k,j}^S,\ \ (-\frac{L_S}{2}<m\leq \frac{L_S}{2},\ k = \frac{2\pi m}{L_S},\ -\frac{M}{2}<N\le\frac{M}{2}).
    \end{aligned}
\eeq
In getting the last equality we used the fact that quasi-momentum $k = \frac{2\pi m}{L_S}$ is equivalent to $k' = \frac{2\pi(m+NL_S)}{L_S}$ for any integer $N$, i.e. $e^{ikj} = e^{i(k+2\pi N)j}$ for integers $j$. 

This result shows that for any eigenstate $\ket{\alpha}$ of the auxiliary system, the common eigenvectors $v_{s,i}$ of $\mathcal{C}_A(\alpha)$ and $\kappa_A(\alpha)$ in (\ref{eq:ckev}) are just the eigenmode wavefunctions of $H_S$, indicating that the entanglement Hamiltonian (\ref{eq:EH_corr}) can be diagonalized simultaneously with $H_S$ and thus $\left[H_S,H_E^A(\alpha)\right] = 0$:
\beq
\begin{aligned}
    &\kappa_{A,g(i)g(j)}(\alpha) = \kappa_{ij}(\alpha) = \sum_{k}\varepsilon_{k}\phi_{k,i}^S\phi_{k,j}^{S*},\ \  \varepsilon_k = \ln\left[\frac{1}{\left(\frac{L_S}{L}\sum_{N}\eta_{\alpha,(m+NL_A)}\right)}-1\right],\\
    &\text{where } k = \frac{2\pi m}{L_S},\ -\frac{L_S}{2}<m\leq\frac{L_S}{2},\\
    &H_S = \sum_{ij\in S}h_{ij}^S c_i^\dag c_j, \ \ h_{ij}^S = \sum_{k}\epsilon_k^S\phi_{k,i}^S\phi_{k,j}^{S*}.
\end{aligned}
\eeq
As discussed in the previous section, this gives rise to the fact that the EHSM method can capture the conserved operators $f_k^\dag f_k$ for all momenta $k$ in the system $S$.

A subtlety in such a degenerate case is that there are other kinds of eigenstates besides the form (\ref{eq:eigenstate}) due to the two-fold degeneracy between modes $k$ and $-k$, since any (normalized) linear combination of $\phi_{k,j}^S$ and $\phi_{-k,j}^S$ is also an eigenmode wavefunction with the same energy. To get all the conserved quantities, we need to include the entanglement Hamiltonian of those eigenstates when constructing EHSM. Explicitly, a many-body eigenstate $\ket{\alpha}$ of the auxiliary system can take a more generic form than \cref{eq:eigenstate}, and to keep it a Fock state for the EHSM method, we take the following form:
\beq
    \ket{\alpha} = \prod_{\widetilde{k}\in \text{1BZ}} \left(\widetilde{f}_{\alpha,\widetilde{k}}^\dagger\right)^{\eta_{\alpha,\widetilde{k}}}\ket{0},
\eeq
where the new set of creation operators $\widetilde{f}_{\alpha,\widetilde{k}}^\dagger$ are defined through a unitary transformation within each single-particle degenerate subspace:
\beq
    \bpm
        \tilde{f}_{\alpha,\widetilde{k}}^\dagger\\
        \tilde{f}_{\alpha,-\widetilde{k}}^\dagger
    \epm
     = U_{\alpha,\widetilde{k}}  
    \bpm
        \tilde{f}_{\widetilde{k}}^\dagger\\
        \tilde{f}_{-\widetilde{k}}^\dagger
    \epm,
    \ \ (0<\widetilde{k}<\pi),\ \ \tilde{f}_{\alpha,0}^\dag = \tilde{f}_{0}^\dag,\ \ \tilde{f}_{\alpha,\pi}^\dag = \tilde{f}_{\pi}^\dag
\eeq
where $U_{\alpha,\widetilde{k}}$ is a $2\times 2$ unitary matrix that is arbitrarily chosen for each momentum $\widetilde{k}>0$. Note that if $L$ is odd, $\tilde{f}_{\pi}^\dag$ does not exist as $\widetilde{k}$ cannot take $\pi$.

Please notice that the subindex $\alpha$ denotes the eigenstate $\ket{\alpha}$ we are constructing, instead of being a new variable. The corresponding eigenmode wavefunctions are also transformed accordingly:
\beq
    \bpm
        \phi_{\alpha,\widetilde{k},j}\\
        \phi_{\alpha,-\widetilde{k},j}
    \epm
     = U_{\alpha,\widetilde{k}}  
    \bpm
        \frac{1}{\sqrt{L}}e^{i\widetilde{k}j}\\
        \frac{1}{\sqrt{L}}e^{i\widetilde{k}j}
    \epm,
    \ \ (0<\widetilde{k}<\pi),
\eeq
for all sites $j\in A\cup B$. It can be seen that a transformed eigenmode $\phi_{\alpha,\widetilde{k},j}$ is a common eigenmode of the eigenmode $\phi_{\alpha,k,j}^S$, $k=M\widetilde{k}\ (\text{mod }2\pi)$ in system $S$, defined through the same unitary transformation:
\beq
    \bpm
        \phi_{\alpha,k,j}^S\\
        \phi_{\alpha,-k,j}^S
    \epm
     = U_{\alpha,\widetilde{k}}  
    \bpm
        \frac{1}{\sqrt{L_S}}e^{ikj}\\
        \frac{1}{\sqrt{L_S}}e^{-ikj}
    \epm,
    \ \ (0<k<\pi),
\eeq
whose corresponding creation operator is also defined through
\beq
    \bpm
        f_{\alpha,k}^\dag\\
        f_{\alpha,-k}^\dagger
    \epm
     = U_{\alpha,\widetilde{k}}  
    \bpm
        f_{k}^\dagger\\
        f_{-k}^\dagger
    \epm,
    \ \ (0<k<\pi).
\eeq
As a result, resembling the derivation in \cref{eq:corr_degenerate}, one gets that the correlation matrix $\mathcal{C}_A(\alpha)$ has eigenvectors $\phi_{\alpha,k,j}^S$, so as the matrix $\kappa(\alpha)$, indicating the corresponding $H_E^A(\alpha)$ is a linear combination of conserved operators $f_{\alpha,k}^\dag f_{\alpha,k}$, $f_0^\dag f_0$, and $f_{\pi}^\dag f_{\pi}$ (if exist). Knowing that 
\beq
    f_{\alpha,k}^\dag f_{\alpha,k} = (u_{11}f_k^\dag + u_{12}f_{-k}^\dag)(u_{11}^*f_k + u_{12}^*f_{-k}) = |u_11|^2 f_k^\dag f_k + |u_{22}|^2 f_{-k}^\dag f_{-k} + u_{11}u_{12}^*f_k^\dag f_{-k} + u_{12}u_{11}^*f_{-k}^\dag f_k,
\eeq
one can see how the EHSM captures the conserved operators $f_k^\dag f_{-k}$ for $0<|k|<\pi$, besides all the $f_k^\dag f_k$.

In addition, we would like to point out that the above discussions can be easily modified for fermion chains with A-PBC: for the integers $n,m$ in quasi-momenta $\widetilde{k}=\frac{2\pi n}{L}$ and $k=\frac{2\pi m}{L_S}$ in the PBC cases, they become half-integers in the A-PBC cases.

\section{Algorithm of evaluating ensemble averages over the non-Abelian GGE}\label{app:GGE_algo}
In this section, we illustrate how to evaluate ensemble averages over the non-Abelian GGE we introduced in the main text. As the only GGE involved in this section is the non-Abelian one, we omit the $({NA})$ superscript throughout this section.

The density matrix of a generic GGE is
\beq
    \rho_{\rm GGE} = \frac{1}{Z}\exp\left(-\sum_{n=1}^{N_Q}\lambda_n Q_S^{(n)}\right)\ ,
\eeq
where the conserved quantities $Q_S^{(n)}$ include $f_k^\dag f_k$ and $f_k^\dag f_{-k}$ for all the quasi-momenta $k$. The problem we are facing now is that these $Q_S^{(n)}$ do not commute with each other, so it is not easy to: (1) fix the Lagrange multipliers $\lambda_n$; (2) evaluate the ensemble average ${\rm Tr}(A\rho_{\rm GGE})$ in the situation that $A$ is a HCB bilinear operator, that is, a linear combination of $a_i^\dag a_j$. 

Let's conquer them one by one. We group the conserved quantities into:
\beq
    \Sigma_\mu^k = \bpm
        f_k^\dagger&f_{-k}^\dagger\epm
        \sigma_\mu
        \bpm f_k\\f_{-k}\epm,\ \ \ \mu = 0,1,2,3,
\eeq
where $k>0$, and $\sigma_\mu$ are the $2\times 2$ Pauli matrices (together with the identity). The only one left is $f_0^\dag f_0$, which can be dealt with separately since it commutes with all other conserved operators. Now, the density matrix becomes
\beq
    \rho_{\rm GGE} = \frac{1}{Z}\exp\left[-\sum_{k>0}\sum_\mu\lambda_\mu^k\Sigma_\mu^k\right]\exp(-\lambda^0 f_0^\dag f_0) = \frac{1}{Z}\left(\prod_{k>0}e^{-\sum_\mu \lambda_\mu^k\Sigma_\mu^k}\right)\exp(-\lambda^0 f_0^\dag f_0).
\eeq
We first evaluate the partition function, which will be helpful later
\beq
    Z={\rm Tr}\left[\left(\prod_{k> 0}e^{-\sum_\mu \lambda_\mu^k\Sigma_\mu^k}\right)\exp(-\lambda^0 f_0^\dag f_0)\right].
\eeq
The difficulty actually arises from the fact that for each $k>0$, the corresponding four operators $\Sigma_\mu^k$ occupy a $2\times 2$ matrix block $\Lambda^k$  in fermion operator basis:
\beq
    \sum_\mu \lambda_\mu^k\Sigma_\mu^k = \bpm
        f_k^\dagger&f_{-k}^\dagger\epm
        \bpm
            \lambda_0^k + \lambda_3^k&\lambda_1^k - i\lambda_2^k\\
            \lambda_1^k + i\lambda_2^k&\lambda_0^k - \lambda_3^k
        \epm
        \bpm f_k\\f_{-k}\epm
        \equiv \bpm
        f_k^\dagger&f_{-k}^\dagger\epm \Lambda^k \bpm f_k\\f_{-k}\epm
\eeq
One way of resolving this problem is to unitarily transform each block $\Lambda^k$ to its diagonal form: $\Lambda^k = V^k D^k V^{k\dag}$, where $D^{k} = diag(d_k, d_{-k})$, $d_{\pm k} = \lambda_0^k\pm\sqrt{\sum_{i=1}^3(\lambda_i^k)^2}$, then one can get the result:
\beq
    \begin{aligned}
    Z &= {\rm Tr}\left[\exp(-\lambda^0 f_0^\dag f_0) \prod_{k > 0}\exp(-d_kf_k^{\text{d}\dagger} f_k^{\text{d}}-d_{-k}f_{-k}^{\text{d}\dagger} f_{-k}^{\text{d}})\right]\\
    &= {\rm Tr}\left[\exp(-\lambda^0 c_0^\dag c_0)\prod_{k\neq 0}\exp(-d_kf_k^{\text{d}\dagger} f_k^{\text{d}})\right]\\
    &= (1+e^{-\lambda^0})\prod_{k\neq 0}(1+e^{-d_k}),
\end{aligned}
\eeq
where the $f_k^{\text{d}}, f_k^{\text{d}\dag}$ operators are defined through $\bpm
        f_k^{\text{d}}\\
        f_{-k}^{\text{d}}
    \epm
     = V^{k\dagger}  
    \bpm
        f_k\\
        f_{-k}
    \epm$, and the superindex $\text{d}$ denotes they being in the basis that diagonalizes the block $\Lambda^k$.They form a complete set of fermionic operators together with $f_0,f_0^\dag$. 

Here we would like to introduce another method that involves an identity for taking trace over the fermionic Fock space:
\beq\label{trace_id}
    {\rm Tr}\left[\exp\left(\sum_{ij}c_i^\dagger X_{ij}c_j\right)\exp\left(\sum_{kl}c_k^\dagger Y_{kl}c_l\right)...\exp\left(\sum_{mn}c_m^\dagger Z_{mn}c_n\right)\right] = {\rm det}\left[I + e^Xe^Y...e^Z\right],
\eeq
which we will be using intensively afterward. The $c,c^\dag$ operators are from a set of fermionic operators obeying the canonical anti-commutation relation, which can represent the $c_j,c_j^\dag$ operators in the real space or the $f_k,f_k^\dag$ operators in the momentum space, and $X,Y,...,Z$ are $L\times L$ matrices. As a result, (a bold $\mathbf{0}$ stands for a $2\times 2$ zero matrix)
\begin{align*}
    Z &={\rm Tr}\left[\exp(-\lambda^0 f_0^\dag f_0) \prod_{k>0}e^{-\sum_\mu\lambda_\mu\Sigma_\mu^k}\right]\\
    &= {\rm det}\left[I+ \exp
    \bpm
        -\lambda^0 & 0 & 0 & ... & 0\\
        0&-\Lambda^{k_1}&\mathbf{0}&...&\mathbf{0}\\
        0&\mathbf{0}&-\Lambda^{k_2}&...&\mathbf{0}\\
        ...\\
        0&\mathbf{0}&\mathbf{0}&...&-\Lambda^{k_{\rm max}}
    \epm
    \right]\\
    &= {\rm det}
    \bpm
        1+e^{-\lambda^0}&0&0&0&...&0\\
        0&1+e^{-d_{k_1}}&0&0&...&0\\
        0&0&1+e^{-d_{-k_1}}&0&...&0\\
        0&0&0&1+e^{-d_{k_2}}&...&0\\
        ...\\
    \epm\\
    &=(1+e^{-\lambda^0})\prod_{k\neq0}(1+e^{-d_k})
\end{align*}
which is exactly the same as the previous result.

Now, we fix the Lagrange multipliers $\lambda$'s by asking the ensemble average of each conserved quantity, ${\rm Tr}(Q_S^{(n)}\rho_{\rm GGE})$ to be the same as its initial value (which is unchanged at a later time due to the conservation) $I_n \equiv \bra{\Psi(t=0)}Q_S^{(n)}\ket{\Psi(t=0)}$. Therefore, we need a closed-form expression for ${\rm Tr}(\Sigma_\mu^k\rho_{\rm GGE})$ and then solve for $\lambda_\mu^k$. (${\rm Tr}(f_0^\dag f_0\rho_{\rm GGE})=1/(e^{\lambda^0}+1)$ is easy to get.)

Recall that in statistical mechanics, one can usually get the values of thermal quantities from the partition function, for example, in a Gibbs (grand canonical) ensemble, we can get the energy $\braket{E}$ and particle number $\braket{N}$ by taking partial derivatives on the partition function $\mathcal{Z}$ with respect to their corresponding Lagrange multipliers $\beta$ (inverse temperature) and $\mu$ (chemical potential):
\beq
\begin{aligned}
    \rho(\beta,\mu) &= \frac{1}{\mathcal{Z}}e^{-\beta(\hat{E}-\mu \hat{N})},\\
    \mathcal{Z} &= {\rm Tr}e^{-\beta(\hat{E}-\mu \hat{N})} = \sum_{j}e^{-\beta(E_j-\mu N_j)},\\
    \Longrightarrow\braket{E}(\beta,\mu)&=-\left(\frac{\partial\ln\mathcal{Z}}{\partial\beta}\right)_{\beta,\mu},\;\;\; 
    \braket{N}(\beta,\mu)=\frac{1}{\beta}\left(\frac{\partial\ln\mathcal{Z}}{\partial\mu}\right)_{\beta,\mu}.
\end{aligned}
\eeq
We can prove that the averages over GGE $\braket{\Sigma_\mu^k}_{\rm GGE}\equiv {\rm Tr}(\Sigma_\mu^k\rho_{\rm GGE})$ can be got in a similar manner: By omitting the summation (production) symbol over $\mu$ and $k \ge 0$ and expanding the exponent, that is, denoting $-\lambda^0f_0^\dag f_0 - \sum_\mu\sum_{k>0}\lambda_\mu^k\Sigma_\mu^k\equiv -\lambda_\mu^k\Sigma_\mu^k$ we get
\beq
    \rho_{\rm GGE} = \frac{1}{Z}\sum_{n=0}^\infty\frac{(-\lambda_\mu^k\Sigma_\mu^k)^n}{n!},
\eeq
then the partial derivative over a particular $-\lambda_\nu^p\;\;(p > 0)$ is
\beq
    \frac{\partial Z\rho_{\rm GGE}}{\partial(-\lambda_\nu^\rho)} = \sum_{n=1}^\infty\frac{1}{n!}\left[\Sigma_\nu^p(-\lambda_\mu^k\Sigma_\mu^k)^{n-1} + (\;\;)\Sigma_\nu^p(\;\;)^{n-2} + ... + (\;\;)^{n-1}\Sigma_\nu^p\right],
\eeq
where all the $(\;\;)$ represent the same thing: $(-\lambda_\mu^k\Sigma_\mu^k)$ with summation over $\mu$ and $k\ge 0$. Here we need to remember that when taking a derivative of a product of some matrices, due to the fact that matrices may not commute to each other, we need to perform the chain rule like above (keeping the order). 

Now, take the trace of the partial derivative:
\begin{align*}
    {\rm Tr}\left[\frac{\partial Z\rho_{\rm GGE}}{\partial(-\lambda_\nu^\rho)}\right] &= \sum_{n=1}^\infty\frac{1}{n!}n{\rm Tr}\left[\Sigma_\nu^p(-\lambda_\mu^k\Sigma_\mu^k)^{n-1}\right]\\
    &= {\rm Tr}\left[\Sigma_\nu^p \left(\sum_{n=1}^\infty \frac{1}{(n-1)!}(-\lambda_\mu^k\Sigma_\mu^k)^{n-1}\right)\right]\\
    &= {\rm Tr}[\Sigma_\nu^p Z\rho_{\rm GGE}] = Z{\rm Tr}[\Sigma_\nu^p\rho_{\rm GGE}],
\end{align*}
and notice that on the left-hand side, the order of taking trace and partial derivative can be exchanged, then
\beq
    \frac{\partial Z {\rm Tr}\rho_{\rm GGE}}{\partial(-\lambda_\nu^p)} = \frac{\partial Z}{\partial (-\lambda_\nu^p)} = Z{\rm Tr}[\Sigma_\nu^p\rho_{\rm GGE}],
\eeq
which can be rewritten into
\beq
    \frac{1}{Z}\frac{\partial Z}{\partial (-\lambda_\nu^p)} = -\frac{\partial}{\partial \lambda_\nu^p}\ln Z = {\rm Tr}[\Sigma_\nu^p\rho_{\rm GGE}]\equiv\braket{\Sigma_\nu^p}_{\rm GGE}.
\eeq
This has exactly the same form as the equilibrium energy from the thermal Gibbs ensemble.

With what we proved just now and the expression of $Z$, we have:
\beq
    \left\{
    \begin{aligned}
    \frac{1}{Z}\frac{\partial Z}{\partial(-\lambda_0^p)} &= \frac{\left[e^{-d_p}(1+e^{-d_{-p}}) + e^{-d_{-p}}(1+e^{-d_{p}})\right]}{(1+e^{-d_p})(1 + e^{-d_{-p}})} = 
    \boxed{\frac{1}{1+e^{d_p}}+\frac{1}{1+e^{d_{-p}}} = \braket{\Sigma_0^p}_{\rm GGE}}\\
    \frac{1}{Z}\frac{\partial Z}{\partial(-\lambda_i^p)} &= \frac{\left[e^{-d_p}(1+e^{-d_{-p}}) - e^{-d_{-p}}(1+e^{-d_{p}})\right]}{(1+e^{-d_p})(1 + e^{-d_{-p}})} \frac{\lambda_i^p}{\sqrt{\sum_{i=1}^3(\lambda_i^p)^2}}\\
    &= \boxed{\left(\frac{1}{1+e^{d_p}}-\frac{1}{1+e^{d_{-p}}}\right)\frac{\lambda_i^p}{\sqrt{\sum_{i=1}^3(\lambda_i^p)^2}} = \braket{\Sigma_i^p}_{\rm GGE}, (i=1,2,3)}
    \end{aligned}
    \right.
\eeq
As a result, the set of equations we have to solve for fixing the four Lagrange multipliers $\lambda_\nu^p$ for each $p>0$ is:
\beq
    \begin{aligned}
&\left\{
    \begin{aligned}
    I_0^p &=\frac{1}{1+e^{d_p}}+\frac{1}{1+e^{d_{-p}}},\\
    I_i^p &= \left(\frac{1}{1+e^{d_p}}-\frac{1}{1+e^{d_{-p}}}\right)\frac{\lambda_i^p}{\sqrt{\sum_{i=1}^3(\lambda_i^p)^2}}, \;\;\;(i=1,2,3)
    \end{aligned}
\right.\\
    &{\rm where}\;\;\; d_{\pm p} = \lambda_0^p\pm\sqrt{\sum_{i=1}^3(\lambda_i^p)^2}.
\end{aligned}
\eeq
This is a set of transcendental equations, but fortunately, they have analytical solutions. We firstly express $d_{\pm p}$ in terms of $I_\mu^p$:
\begin{align*}
    I_0^p +\sqrt{\sum_{i=1}^3(I_i^p)^2} = \frac{2}{1+e^{d_p}}\Longrightarrow d_p = \ln\left(\frac{2}{I_0^p +\sqrt{\sum_{i=1}^3(I_i^p)^2}}-1\right),\\
    I_0^p -\sqrt{\sum_{i=1}^3(I_i^p)^2} = \frac{2}{1+e^{d_{-p}}}\Longrightarrow d_{-p} = \ln\left(\frac{2}{I_0^p -\sqrt{\sum_{i=1}^3(I_i^p)^2}}-1\right),
\end{align*}
and then the Lagrange multipliers can be expressed as:
\begin{align}
    \lambda_0^p &= (d_p + d_{-p})/2,\\
    \sqrt{\sum_{i=1}^3(\lambda_i^p)^2} &= (d_p - d_{-p})/2,\nonumber\\
    \Longrightarrow\lambda_i^p &= \frac{I_i^p(d_p - d_{-p})/2}{\left(\frac{1}{1+e^{d_p}}-\frac{1}{1+e^{d_{-p}}}\right)} = \frac{I_i^p(1+e^{d_p})(1+e^{d_{-p}})(d_p-d_{-p})}{2(e^{d_{-p}} - e^{d_p})}.
\end{align}
These fix the density matrix $\rho_{\rm GGE}$.

Finally, we come to the second part of our goal: evaluate $\braket{a_i^\dagger a_j}_{\rm GGE} = {\rm Tr}\left[a_i^\dagger a_j\rho\right]$ in order to calculate ensemble average of any bosonic bilinear operator. This part follows the same logic as in \cite{GGE_algo}.

Let's take $i<j$ for the moment ($i>j$ can be derived likewise, and we will take care of $i=j$ later). We will be intensively using the identity (\ref{trace_id}). To be convenient, we rewrite the exponent part of the density matrix as
\beq
    \rho_{\rm GGE} = \frac{1}{Z}\exp(-\lambda_\mu^k\Sigma_\mu^k)= \frac{1}{Z}\exp(-\sum_{m,n}c_m^\dagger\Lambda_{mn}c_n),
\eeq
where $m,n$ represent lattice sites and the matrix $\Lambda$ is the $L\times L$ block diagonal matrix with $2\times2$ matrices $\Lambda^k$ on the diagonal while \textbf{transformed back into real space basis}. 

Now we evaluate the trace
\begin{align*}
    {\rm Tr}\left[a_i^\dagger a_j\rho_{\rm GGE}\right] &= \frac{1}{Z}{\rm Tr}\left[a_i^\dagger a_j\exp(-\sum_{m,n}c_m^\dagger\Lambda_{mn}c_n)\right]\\
    &= \frac{1}{Z}{\rm Tr}\left[\prod_{\delta=1}^{i-1}\left(1-2c_\delta^\dagger c_\delta\right)c_i^\dagger c_j\prod_{\gamma=1}^{j-1}\left(1-2c_\gamma^\dagger c_\gamma\right)\exp(-\sum_{m,n}c_m^\dagger\Lambda_{mn}c_n)\right]\\
    &= \frac{1}{Z}{\rm Tr}\left[c_i^\dagger c_j\prod_{\gamma=1}^{j-1}\left(1-2c_\gamma^\dagger c_\gamma\right)\exp(-\sum_{m,n}c_m^\dagger\Lambda_{mn}c_n)\prod_{\delta=1}^{i-1}\left(1-2c_\delta^\dagger c_\delta\right)\right]\\
    &= \frac{1}{Z}{\rm Tr}\left[c_i^\dagger c_j\prod_{\gamma=1}^{j-1}\exp(i\pi c_\gamma^\dagger c_\gamma)\exp(-\sum_{m,n}c_m^\dagger\Lambda_{mn}c_n)\prod_{\delta=1}^{i-1}\exp(i\pi c_\delta^\dagger c_\delta)\right].
\end{align*}
The factors in the trace can be expressed in the form of $\exp\left(\sum_{ij}c_i^\dag X_{ij}c_j\right)$, by noticing that
\beq
    c_i^\dagger c_j = \exp\left(\sum_{mn}c_m^\dagger A_{mn} c_n\right)-1,
\eeq
where the only nonzero element of matrix $A$ is $A_{ij} = 1$, and that
\beq
    \prod_{\gamma=1}^{j-1}\exp(i\pi c_\gamma^\dagger c_\gamma) = \exp\left[
    \bpm
        c_1^\dagger&c_2^\dagger&...&c_L^\dagger
    \epm
    \bpm
        i\pi&0&0&0&0&0&...&0\\
        0&i\pi&0&0&0&0&...&0\\
        ...\\
        0&0&...&0&i\pi&0&...&0\\
        0&...&...&...&...&...&...&0\\
        ...\\
        0&...&...&...&...&...&...&0
    \epm
    \bpm
        c_1\\c_2\\...\\c_L
    \epm
    \right],
\eeq
where the $L\times L$ diagonal matrix in the middle, named $A_1$, has its first $j-1$ diagonal elements to be $i\pi$ and the rest of them to be $0$. (A similar formula can be found for the other product while the matrix in the middle with first $i-1$ diagonal elements being $i\pi$ is called $A_2$.)

Now, with the help of (\ref{trace_id}), we simplify the trace over the huge many-body Hilbert space to the evaluation of determinants of $L\times L$ matrices (here $I$ represents $L\times L$ identity matrix):
\beq
    {\rm Tr}\left[a_i^\dagger a_j\rho_{\rm GGE}\right] = \frac{1}{Z}\left\{{\rm det}\left[I+e^A e^{A_1} e^{-\Lambda} e^{A_2}\right] - {\rm det}\left[I+ e^{A_1} e^{-\Lambda} e^{A_2}\right]\right\}.
\eeq
Evaluating the exponents is a simple task:
\begin{align*}
    e^A &= I + A + \frac{A^2}{2} + ... = I + A,\\
    e^{A_1} &=     
    \bpm
        e^{i\pi}&0&0&0&0&0&...&0\\
        0&e^{i\pi}&0&0&0&0&...&0\\
        ...\\
        0&0&...&0&e^{i\pi}&0&...&0\\
        0&...&...&...&...&...&...&0\\
        ...\\
        0&...&...&...&...&...&...&0
    \epm
     = 
    \bpm
        -1&0&0&0&0&0&...&0\\
        0&-1&0&0&0&0&...&0\\
        ...\\
        0&0&...&0&-1&0&...&0\\
        0&...&...&...&...&...&...&0\\
        ...\\
        0&...&...&...&...&...&...&0
    \epm\equiv O_1,\\
    e^{A_2}&\equiv O_2.
\end{align*}
Therefore, 
\beq
    {\rm Tr}\left[a_i^\dagger a_j\rho_{\rm GGE}\right] = \frac{1}{Z}\left\{{\rm det}\left[I+\left(I+A\right) O_1 e^{-\Lambda} O_2\right] - {\rm det}\left[I+ O_1 e^{-\Lambda} O_2\right]\right\}.
\eeq
This result, as one can convince themself, also applies to the situation that $i>j$. With these two-point functions evaluated, we will be able to calculate the ensemble averages of a variety of operators that can be easily expressed as products of the bosonic operators.

For the special case that $i=j$, one can easily show by definition that $a_j^\dag a_j = c_j^\dag c_j$, so the problem of evaluating ${\rm Tr}\left[a_j^\dagger a_j\rho_{\rm GGE}\right]$ is covered the evaluation of the fermionic two-point functions: ${\rm Tr}\left[c_i^\dagger c_j\rho\right]$ or in momentum space ${\rm Tr}\left[f_k^\dagger f_{k'}\rho\right]$. This can be done very similarly to the bosonic two-point functions:
\beq
\begin{aligned}
    {\rm Tr}\left[c_i^\dagger c_j\rho\right] &= \frac{1}{Z}{\rm Tr}\left[c_i^\dagger c_j\exp(-\sum_{m,n}c_m^\dagger\Lambda_{mn}c_n)\right]\\
    &= \frac{1}{Z}{\rm Tr}\left[\left(\exp\left(\sum_{mn}c_m^\dagger A_{mn} c_n\right)-1\right)\exp(-\sum_{m,n}c_m^\dagger\Lambda_{mn}c_n)\right]\\
    &= \frac{1}{Z}\left\{{\rm det}\left[I+(I+A)e^{-\Lambda}\right] - {\rm det}\left[I + e^{-\Lambda}\right]\right\}.
\end{aligned}
\eeq

Lastly, by Fourier transform of $\braket{a_i^\dagger a_j}_{\rm GGE}$, one can obtain the momentum space bosonic bilinears $\braket{b_k^\dagger b_{\pm k}}_{\rm GGE}$, which are shown in the main text.

\section{Additional numerical results of GGE in homogeneous free-fermion lattice}\label{app:GGE_add}
Here we present some additional numerical results about the GGE. For the PBC case, we showed the relaxation results of HCB observables $b_k^\dagger b_k$ and the imaginary part of non-Hermitian operators $b_k^\dagger b_{-k}$, comparing with the prediction given by Abelian and non-Abelian GGEs. Here we show the result for the real part of $b_k^\dag b_{-k}$, see Fig. \ref{fig4}.
\begin{figure}[h!]
    \if\flag1\includegraphics{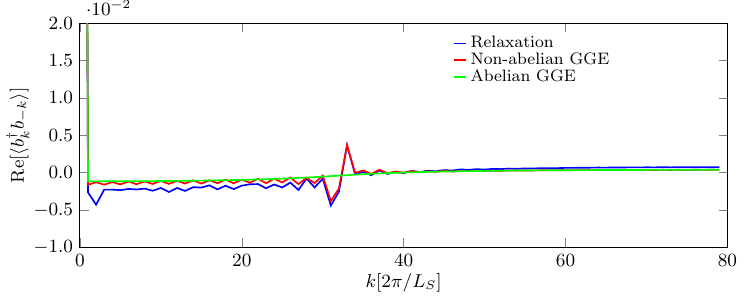}\\\else\include{fig4}\fi
    \caption{Relaxed values of the real part of $\braket{b_k^\dag b_{-k}}$ predicted by the Abelian GGE and non-Abelian GGE, compared with the results from time-evolution calculations.}
\end{figure}
It can be seen that the difference among the three curves is smaller, but the non-Abelian GGE still captures more features of the curve representing relaxed values. In addition, we show in \cref{fignn} that the non-Abelian GGE provides better predictions on the correlators $\braket{n_i n_j}=\braket{c_i^\dag c_i c_j^\dag c_j}$ than the Abelian GGE, giving a piece of evidence that the non-Abelian GGE also improves the predictions on two-body fermion observables.

\begin{figure}
\centering
\begin{tabular}{ l l l }
    (a)&\ &(b)\\
    \includegraphics[width=0.5\columnwidth]{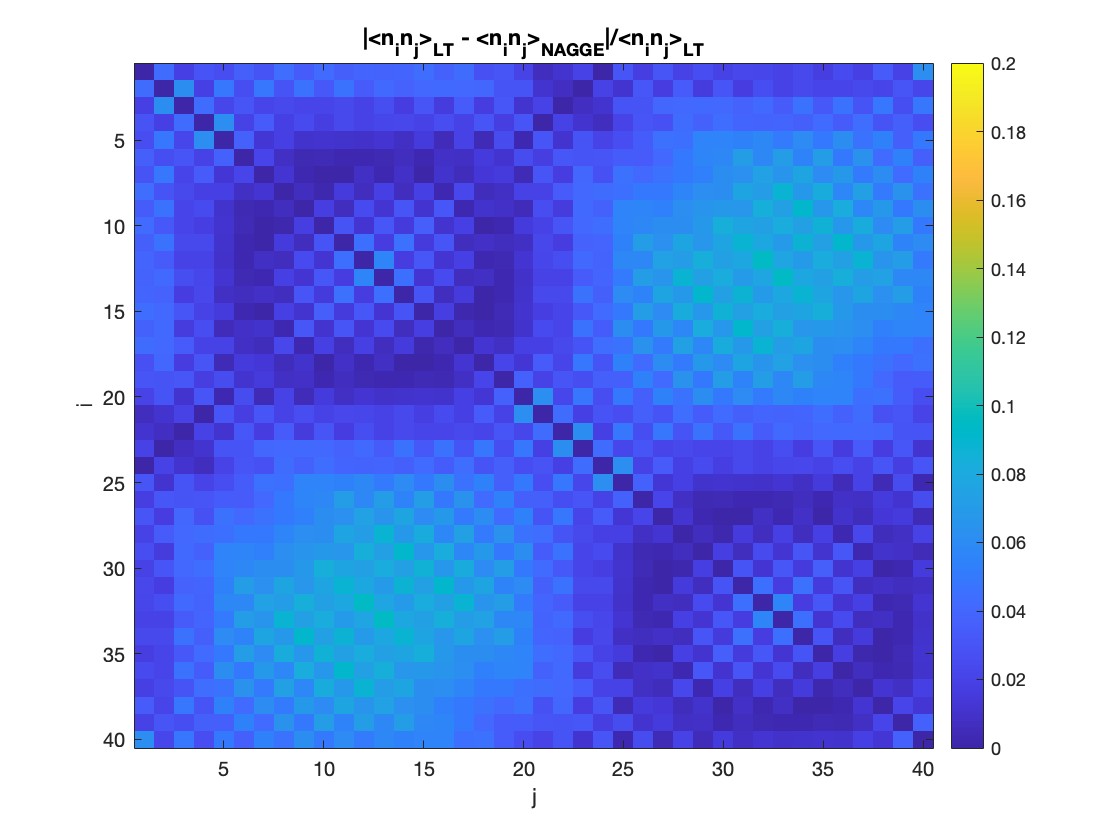}&\ &\includegraphics[width=0.5\columnwidth]{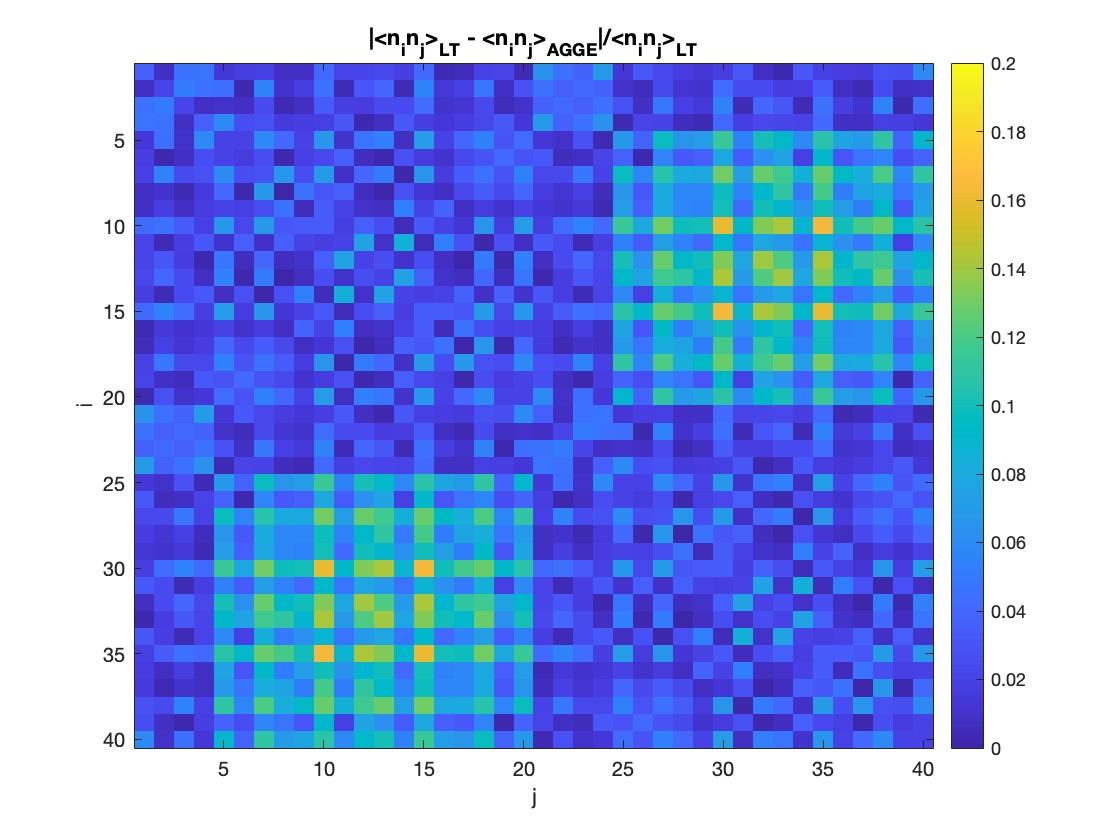}\\
\end{tabular}
\caption{Relative error of the correlator $\braket{n_i n_j}$ between the long-time average $\braket{n_in_j}_{\rm LT}$ and (a) the predictions of the non-Abelian GGE $\braket{n_in_j}_{\rm NAGGE}$ (b) the predictions of the Abelian GGE $\braket{n_in_j}_{\rm AGGE}$. } 
\label{fignn}
\end{figure}

\begin{figure}
    \centering
    \if\flag1\includegraphics{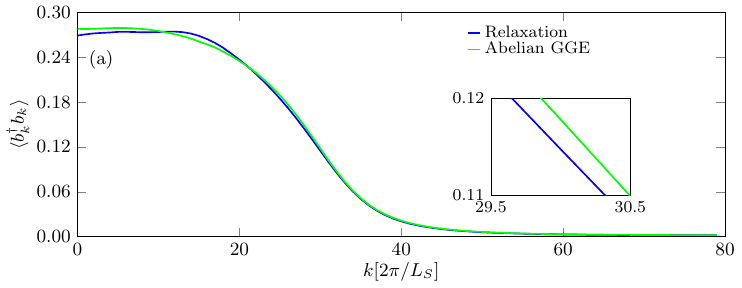}\else\include{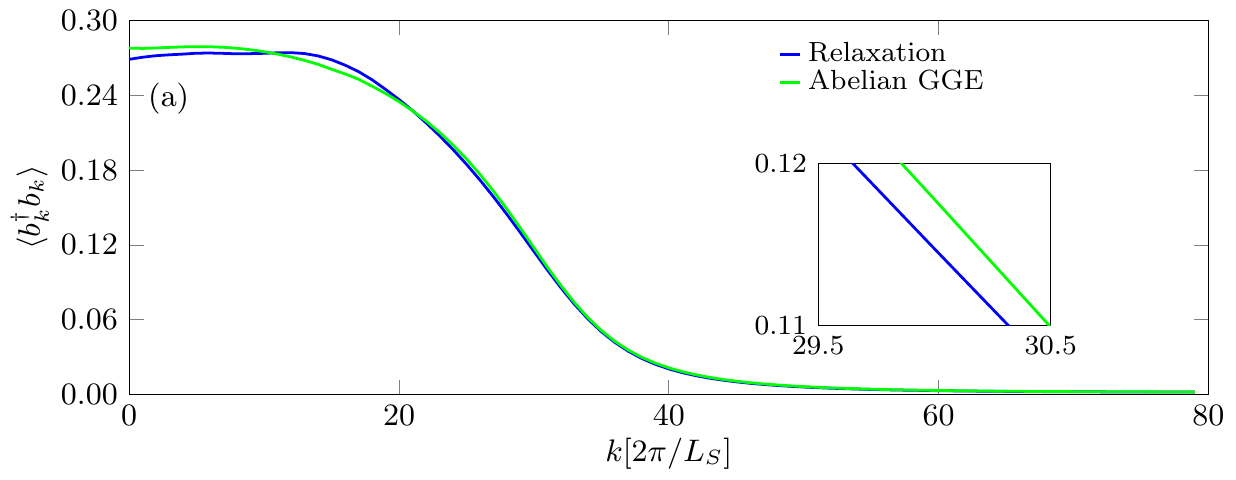}\fi\\
    \if\flag1\includegraphics{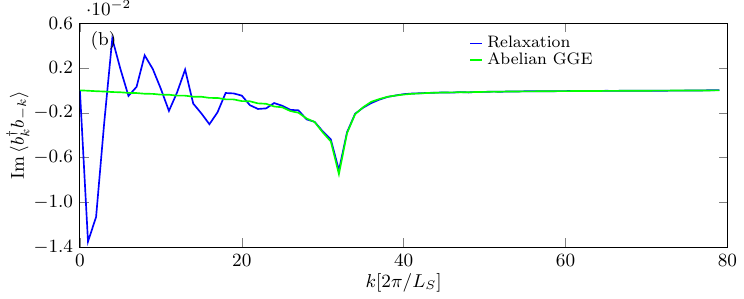}\else\include{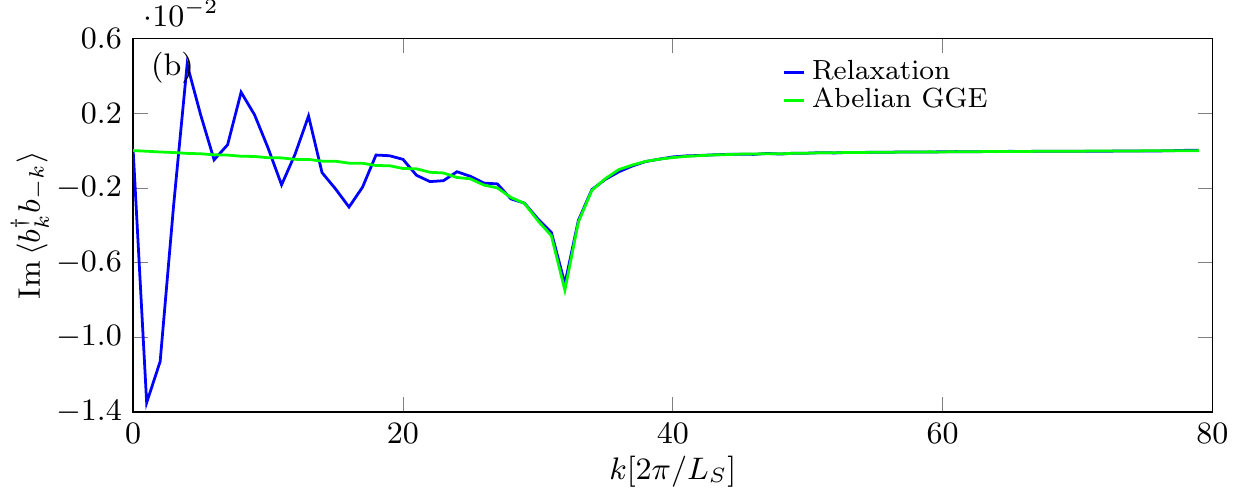}\fi\\
    \if\flag1\includegraphics{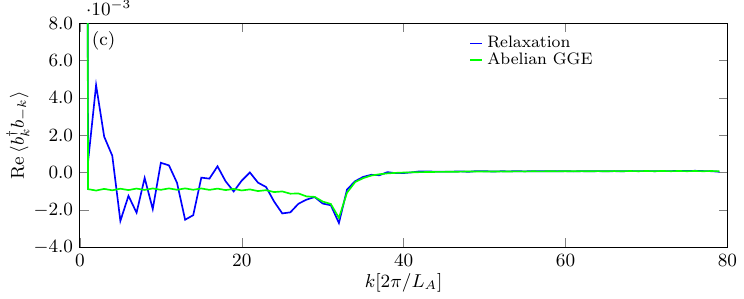}\else\include{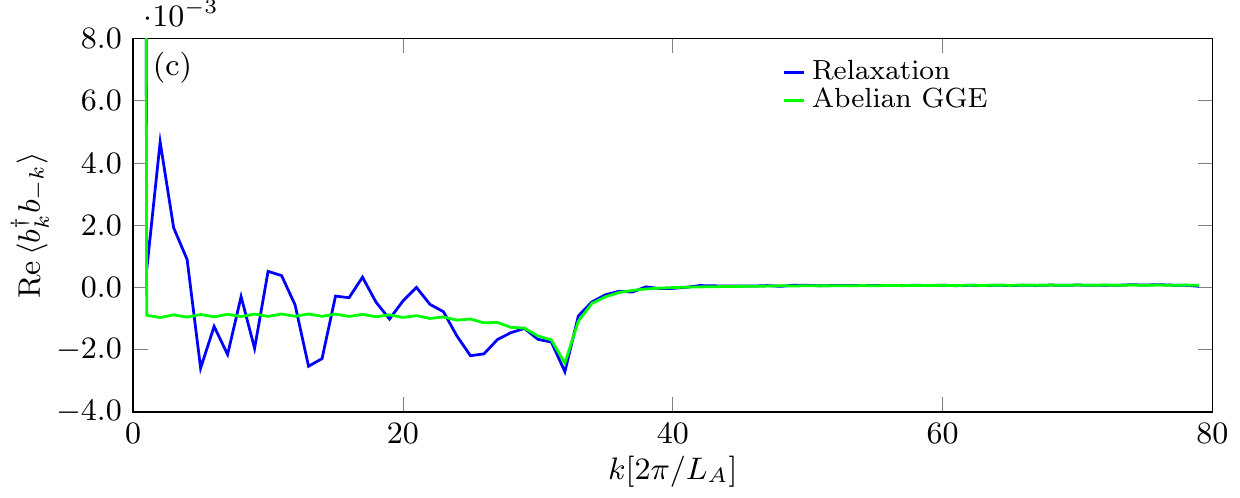}\fi
    \caption{Relaxed values of hard-core boson operators $\braket{b_k^\dag b_k}$ and $\braket{b_k^\dag b_{-k}}$ in a homogeneous OBC system predicted by the GGE, compared with the results from time-evolution calculations.}
    \label{fig5}
\end{figure}

For the homogeneous OBC case, we only get all the mutually commuting mode occupation operators as conserved quantities from entanglement Hamiltonian, then the corresponding GGE is still Abelian. The numerical results are in Fig. \ref{fig5}.

\section{Numerical results of EHSM in a generic free-fermion lattice with OBC}
\begin{figure}
\centering
\begin{tabular}{ l l l }
    (a)&\ &(b)\\
    \if\flag1\includegraphics[width=0.4\columnwidth]{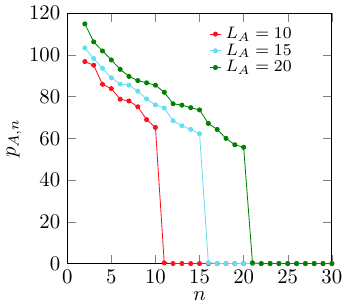}\else\include{essay_homo/figurefix}\fi&\ &\includegraphics[width=0.5\columnwidth]{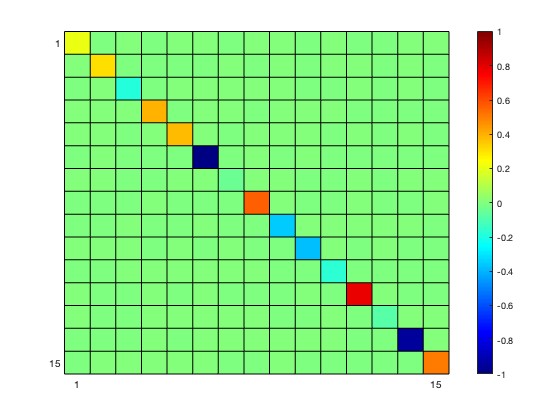}\\
\end{tabular}
\caption{(a)The EHSM eigenvalues for generic free fermion chains with OBC. $L_A = 10,15,20$ and $L = 2L_A+1$. (b) The matrix elements of a typical EHSM eigen-operator in the eigenbasis of the corresponding $H_S$.}
\label{fig6}
\end{figure}

In order to show that the way we proposed of designing an auxiliary system for any arbitrary free fermion chain indeed works, we show the numerical results of EHSM eigenvalues and a typical EHSM $\kappa_A$ matrix in eigenmode space. As mentioned in the main text, the size of the auxiliary system is $L = 2L_S+1$, and the Hamiltonian of the auxiliary system is designed to be mirror symmetric about site $1\leq j\leq S$ and $h_{ij} = h_{ij}^{S}$ if $i,j\leq L_S$. See Fig. \ref{fig6}. 

\section{Proof: the non-Abelian GGE correctly captures the long-time averages of all fermionic biliears}
In the main text, we claimed that in a homogeneous 1D free fermion lattice with periodic boundary conditions, the long-time averages of fermion bilinears: $\overline{\braket{c_i^\dag c_j}}$, where $i,j$ are indices of sites, is accurately predicted by the ensemble average of the non-Abelian GGE: ${\rm Tr}[\rho_{\rm GGE}^{(NA)}c_i^\dag c_j]$. Here we give a proof of it.

In this model, the creation and annihilation operators on sites, $c_j^\dag, c_j$ can be expressed by the eigenmode operators $f_k^\dag, f_{k}$ by
\beq
    c_j^\dag = \sum_k \frac{e^{-ikj}}{\sqrt{L_S}}f_k^\dag,\quad c_j = \sum_k\frac{e^{ikj}}{\sqrt{L_S}} f_k,
\eeq
therefore,
\beq
    c_i^\dag c_j = \sum_{kk'}\frac{e^{i(k'j-ki)}}{L_S}f_k^\dag f_{k'}.
\eeq
The fermion bilinears are then $\braket{c_i^\dag c_j}(t) = \bra{\Psi(t)}c_i^\dag c_j\ket{\Psi(t)}$, where $\ket{\Psi(t)} = e^{-iH_S t}\ket{\Psi}$ is the quantum state of the system at time $t$ evolved from $\ket{\Psi}$ at $t=0$, then
\beq
    \braket{c_i^\dag c_j}(t) = \sum_{kk'}\frac{e^{i(k'j-ki)}}{L_S}\braket{f_k^\dag f_{k'}}(t).
\eeq
Simple calculation shows that $\braket{f_k^\dag f_{k'}}(t) = e^{i(\epsilon_k-\epsilon_{k'})t}\braket{f_k^\dag f_{k'}}$, where $\braket{f_k^\dag f_{k'}} = \bra{\Psi}f_k^\dag f_{k'}\ket{\Psi}$. As a result, when taking long-time averages, only the terms with $\epsilon_k-\epsilon_{k'} = 0$ survive, i.e., the terms involving $\braket{f_k^\dag f_k}$ and $\braket{f_k^\dag f_{-k}}$
\beq
    \overline{\braket{c_i^\dag c_j}} = \sum_k\frac{e^{ik(j-i)}}{L_S}\braket{f_k^\dag f_k} + \sum_{k\neq 0,\pi}\frac{e^{ik(j+i)}}{L_S}\braket{f_k^\dag f_{-k}}.
\eeq

Similarly, for the ensemble average ${\rm Tr}[\rho_{\rm GGE}^{(NA)}c_i^\dag c_j]$, one has
\beq
    {\rm Tr}[\rho_{\rm GGE}^{(NA)}c_i^\dag c_j] = \sum_{kk'}\frac{e^{i(k'j-ki)}}{L_S}{\rm Tr}[\rho_{\rm GGE}^{(NA)}f_k^\dag f_{k'}].
\eeq
Since $\rho_{\rm GGE}^{(NA)} = (\frac{1}{Z})e^{-\sum_k\lambda_k^{(+)}f_k^\dag f_k + \lambda_k^{(-)}f_k^\dag f_{-k}}$, where $\lambda_{k}^{(\pm)}$ are fixed by ${\rm Tr}[\rho_{\rm GGE}^{(NA)}f_k^\dag f_{\pm k}] = \braket{f_k^\dag f_{\pm k}}$, it can be shown that for $f_k^\dag f_{k'}$ such that $k'\neq \pm k$, $[\rho_{\rm GGE}^{(NA)}f_k^\dag f_{k'}] = 0$. As a result, 
\beq
\begin{aligned}
    {\rm Tr}[\rho_{\rm GGE}^{(NA)}c_i^\dag c_j] &= \sum_{k}\frac{e^{ik(j-i)}}{L_S}{\rm Tr}[\rho_{\rm GGE}^{(NA)}f_k^\dag f_{k}] + \sum_{k\neq 0,\pi}\frac{e^{ik(j+i)}}{L_S}{\rm Tr}[\rho_{\rm GGE}^{(NA)}f_k^\dag f_{-k}]\\
    &= \sum_k\frac{e^{ik(j-i)}}{L_S}\braket{f_k^\dag f_k} + \sum_{k\neq 0,\pi}\frac{e^{ik(j+i)}}{L_S}\braket{f_k^\dag f_{-k}}\\
    &= \overline{\braket{c_i^\dag c_j}}.
\end{aligned}
\eeq

\end{document}